\documentstyle[prb,aps,eqsecnum,epsf,multicol]{revtex}

\newcommand{\bleq}{\ifpreprintsty
                   \else
                   \end{multicols}\vspace*{-3.5ex}{\tiny
                   \noindent\begin{tabular}[t]{c|}
                   \parbox{0.493\hsize}{~} \\ \hline \end{tabular}}
                   \fi}
\newcommand{\eleq}{\ifpreprintsty
                   \else
                   {\tiny\hspace*{\fill}\begin{tabular}[t]{|c}\hline
                    \parbox{0.49\hsize}{~} \\
                    \end{tabular}}\vspace*{-2.5ex}\begin{multicols}{2}
                    \fi}
\newcommand{\bcols}{\ifpreprintsty\else\begin{multicols}{2}\fi}
\newcommand{\ecols}{\ifpreprintsty\else\end{multicols}\fi}

\begin{document} 
\draft
\title{Effect of umklapp scattering on the magnetic-field-induced
spin-density waves 
in quasi-one-dimensional organic conductors  } 
\author{ N. Dupuis\cite{leave} and V.M. Yakovenko }
\address{Department of Physics, University of Maryland, College Park, MD
20742-4111} 
\date{March 6, 1998} 
\maketitle
\begin{abstract}
We study the effect of umklapp scattering on the magnetic-field-induced
spin-density-wave (FISDW) phases which are experimentally observed in the
quasi-one-dimensional organic conductors of the Bechgaard salts family. 
Within the framework of the quantized nesting model, we show that the
transition temperature is determined by a modified Stoner 
criterion which includes the effect of umklapp scattering. 
We determine the SDW polarization (linear or circular) by analyzing the
Ginzburg-Landau expansion of the free energy. We also 
study how umklapp processes modify the quantum Hall effect (QHE) and the
spectrum of the FISDW phases. We find that 
umklapp scattering stabilizes phases which exhibit a sign reversal of the
QHE, as experimentally observed in the Bechgaard
salts. These ``negative'' phases are characterized by the simultaneous
existence of two SDWs with comparable amplitudes. As the umklapp scattering
strength increases, they may become helicoidal (circularly
polarized SDWs). The QHE vanishes in the helicoidal phases, but a
magnetoelectric effect appears. These two characteristic properties may be
utilized to detect the magnetic-field-induced helicoidal SDW phases
experimentally. 
\end{abstract}

\bcols
\section{Introduction}

The organic conductors of the Bechgaard salts family (TMTSF)$_2$X (where
TMTSF stands for tetramethyltetraselenafulvalene) exhibit a rich phase
diagram when temperature, magnetic field, or pressure are varied.  In
three members of this family (X=ClO$_4$, PF$_6$, ReO$_4$), a moderate
magnetic field above several Tesla destroys the metallic phase and induces
a series of SDW phases separated by first-order
phase transitions.\cite{rev} Because of a strong quasi-one-dimensional
anisotropy (the typical ratio of the electron transfer integrals in
the three crystal directions is $t_a:t_b:t_c=3000:300:10$ K), the
Fermi surfaces of these materials are open.  According to the
so-called quantized nesting model (QNM),
\cite{rev,Gorkov84,Heritier84,Montambaux85,Yamaji85,Virosztek86,Montambaux88} 
the formation of the FISDW phases results from an 
interplay between the nesting
properties of the Fermi surface and the quantization of the electronic
orbits in magnetic field. The wave vector of
a FISDW adjusts itself to the magnetic field so that unpaired electrons
completely fill an integer number of Landau levels, thus the Hall
effect is quantized.\cite{Poilblanc87,Yakovenko91} The standard QNM
predicts the Hall plateaus of the same
sign, referred to as positive by convention, which agrees with most
experiments. However, at certain pressures, a negative Hall effect
is also observed.\cite{Ribault85,Piveteau86,Hannahs89,Cooper89,Balicas95} 

We have recently shown that umklapp processes may naturally explain the sign
reversals of the QHE. \cite{ND98} Our explanation differs from the
one suggested by Zanchi and Montambaux \cite{Zanchi96} in invoking the
pressure dependence of umklapp scattering rather than the electron band
structure. Although both explanations lead to similar phase diagrams, we
predict the existence of two SDWs (with comparable amplitudes) in the
negative phases. Moreover, we have shown that the negative phases are
likely to become helicoidal (circularly polarized SDWs) under the effect of
stronger umklapp scattering. Experimentally, this can be achieved by
decreasing the applied pressure. The helicoidal phases are characterized by
a vanishing QHE and a kinetic magnetoelectric effect. \cite{ND98} 

In this paper we study the effect of umklapp processes on the
FISDW phases within the framework of the QNM. We discuss in more detail the
results reported in Ref.\ \onlinecite{ND98}, and address issues not
considered in the latter. The effect
of umklapp scattering on the FISDW phases was studied before by Lebed'
using rather crude approximations, but the helicoidal phases and the sign
reversals of the QHE were not discussed.\cite{Lebed90,Lebed91} 

In Bechgaard salts, complete charge transfer from the
molecules TMTSF	to the anions X leads to a conduction band that is
quarter-filled in terms of holes. A dimerization along the $x$ axis
induces a gap in the electronic spectrum. This results in a half-filled band
for the holes so that umklapp processes transfering $4k_F=2\pi /a$ are
possible ($k_F$ being the Fermi wave vector of the holes, and $a$ the
lattice spacing along the chains). Therefore, a quasi-1D g-ology
description of the FISDW phases should include not only forward ($g_2$) and
backward ($g_1$) scattering amplitudes, but also umklapp scattering
amplitude ($g_3$). \cite{Barisic81,Emery82} Since the
dimerization is weak, we expect the umklapp scattering amplitude $g_3$ to be
small. Nevertheless, we shall show in this
paper that very weak umklapp processes can have drastic effects on the
low-temperature phase diagram.

In the next section we obtain the FISDW transition temperature in the random
phase approximation (RPA) and discuss the phase diagram. In absence of
umklapp scattering, the 
instability of the metallic phase corresponds to the formation
of a SDW with a quantized longitudinal wave vector $Q_x^{(N)}=2k_F+NG$
($N$ integer) where $G=eHb/\hbar $ is a magnetic wave vector (with $H$ the
magnetic field, $b$ the interchain spacing, and $-e$ the electron charge). 
As the field varies, the value of the integer $N$ changes (its sign
remaining the same), which leads to a cascade of FISDW phases 
separated by first order transitions. The integer $N$ also determines the
quantum Hall conductivity: $\sigma _{xy}=-2Ne^2/h$ per one layer of the
TMTSF molecules. \cite{rev} In presence of umklapp
scattering, two SDWs with quantized longitudinal wave vectors $Q_x^{(N)}$
and $Q_x^{(-N)}$ form simultaneously. We label each FISDW phase by the
integer $N$ such that the SDW with wave vector $Q_x^{(N)}$ has the largest
amplitude. We find that 
the transition temperature $T_c^{(N)}$ is determined by a modified Stoner
criterion that includes the effect of umklapp scattering. We calculate
$T_c^{(N)}$ numerically as a function of $g_3$ and the detailed geometry of
the Fermi surface. In order to keep our discussion of the phase
diagram compact, we use some results that are proved in subsequent sections. 
We find that weak umklapp scattering ($g_3/g_2$ a few percents) can lead to
a FISDW cascade with both positive and negative values of $N$. Since the
quantum Hall conductivity is still determined by the integer $N$, $\sigma
_{xy}=-2Ne^2/h$ (section \ref{sec:qhe}), this leads to
sign reversals of the QHE as the magnetic field varies. The negative
phases (i.e., with a sign reversed QHE) correspond to even integers $N$ in
agreement with experiments. For stronger values of $g_3$, the negative
phases may become helicoidal (i.e., with a circular polarization of the
SDWs) (section \ref{sec:polar}). The helicoidal phases are characterized by
a vanishing QHE and a kinetic magnetoelectric effect (sections \ref{sec:qhe}
and \ref{sec:mee}). There may also be some reentrances of the phase $N=0$
within the cascade. In section \ref{sec:spec}, we study how umklapp
scattering affects the excitation spectrum in the FISDW phases. In section
\ref{sec:coex}, we consider the possible coexistence of two successive FISDW
phases, an issue previously considered by Lebed' in the case of sinusoidal
SDWs.\cite{Lebed90} 

The experimental consequences of our work are mainly discussed in sections
\ref{sec:imp}, \ref{sec:mee} and \ref{sec:sum}. To a large extent, these
sections can be read independly of the rest of the paper.

\section{Instability of the metallic phase and phase diagram }
\label{sec:imp}

In this section, we first consider the system with no electron-electron
interaction. We obtain the one-particle eigenstates in the
presence of a uniform magnetic field $H$ along the least conducting axis $z$, 
and calculate the bare susceptibility. Then we take into account the
interactions (including umklapp processes) and study the formation of SDWs
in the RPA. We also discuss the QHE and the polarization of the SDWs using
results to be proved in the next sections. 

In the vicinity of the Fermi energy, the electron dispersion law in
the Bechgaard salts is approximated as (we take $\hbar =k_B=1$ throughout
the paper, and the Fermi energy $\mu $ is chosen as the origin of the energies)
\begin{equation} 
  E(k_x,k_y) = v_F(|k_x| -k_F) + t_\perp(k_yb),
\label{disp}
\end{equation}
where $k_x$ and $k_y$ are the electron momenta along and across the
one-dimensional chains of TMTSF. In Eq.\ (\ref{disp}),
the longitudinal electron dispersion is linearized in $k_x$ in the
vicinity of the two one-dimensional Fermi points $\pm k_F$, and
$v_F=2at_a\sin(k_Fa)$ is the corresponding Fermi velocity. The function
$t_\perp (u)$, which describes the
propagation in the transverse direction, is periodic: $t_\perp
(u)=t_\perp (u+2\pi )$. It can be expanded in Fourier series
\begin{eqnarray}
t_\perp (u) &=& -2t_b \cos (u) -2t_{2b} \cos (2u) \nonumber \\ 
&& -2t_{3b}\cos (3u) -2t_{4b}\cos (4u) \cdots 
\end{eqnarray}
If we retain only the first harmonic ($t_b$), we obtain a Fermi surface with
perfect nesting at $(2k_F,\pi /b)$. The other harmonics $t_{2b},t_{3b}\cdots
\ll t_b$ generate deviations from the perfect nesting. They
have been introduced in order to keep a realistic description of the Fermi
surface despite the linearization around $\pm k_F$. \cite{rev} In the
following, we shall retain only $t_b$, $t_{2b}$, and $t_{4b}$ (as we shall
show, $t_{3b}$ does not play an important role and can be discarded). We do
not consider the electron dispersion along the $z$ axis, because it is not
important in the following (its main effect is to introduce a 3D threshold
field below which the FISDW cascade is suppressed\cite{rev}).  

The effect of the magnetic field along the $z$ axis is taken into account
via the Peierls  
substitution ${\bf k}\to -i\bbox{ \nabla }-e{\bf A}$. (The charge
$e$ is positive since the actual carriers are holes.) Following Ref.\
\onlinecite{ND95}, we use the gauge ${\bf A}=(-Hy,0,0)$. Considering also
electron-electron interactions, we obtain the Hamiltonian ${\cal 
H}={\cal H}_0+{\cal H}_{\rm int}$ with 
\begin{eqnarray}
{\cal H}_0 &=& \sum _{\alpha ,\sigma } \int d^2 r 
\psi _{\alpha \sigma }^\dagger ({\bf r}) \lbrack v_F(-i\alpha \partial _x
-k_F) +\alpha \hat m \omega _c \nonumber \\ && 
+t_\perp  (-ib\partial _y) +\sigma \mu _B H 
\rbrack \psi _{\alpha \sigma } ({\bf r}), \nonumber \\ 
{\cal H}_{\rm int} &=& \frac{g_2}{2} \sum _{\alpha ,\sigma ,\sigma '} 
 \int d^2 r 
\psi _{\alpha \sigma }^\dagger ({\bf r})
\psi _{\bar \alpha \sigma '}^\dagger ({\bf r})
\psi _{\bar \alpha \sigma '}({\bf r})
\psi _{\alpha \sigma } ({\bf r}) \nonumber \\
&& + \frac{g_3}{2} \sum _{\alpha ,\sigma } 
 \int d^2 r e^{-i\alpha 4k_Fx} 
\nonumber \\ && \times 
\psi _{\bar \alpha \sigma }^\dagger ({\bf r})
\psi _{\bar \alpha \bar \sigma }^\dagger ({\bf r})
\psi _{\alpha \bar \sigma } ({\bf r})
\psi _{\alpha \sigma } ({\bf r}) . 
\label{ham1}
\end{eqnarray} 
Here $\psi _{\alpha \sigma } ({\bf r})$ are fermionic operators for
right ($\alpha =+$) and left ($\alpha =-$) moving particles. $\sigma =+(-)$
for $\uparrow $ ($\downarrow $) spin. We use the notation ${\bf r}=(x,mb)$
($m$ integer), $\int d^2r =b\sum _m \int dx$ and $\bar \alpha =- \alpha $,
$\bar \sigma =-\sigma $.

Apart from the Zeeman term $\sigma \mu _BH$ ($\mu _B$ is the Bohr magneton,
and we take the electron gyromagnetic factor $g$ equal to
2), the magnetic field introduces the additional term $\alpha \hat m \omega
_c$ where $\hat m$ is the (discrete) position operator in the $y$
direction. $\omega _c=Gv_F=eHbv_F$. 

The  interacting part of the Hamiltonian contains two terms corresponding to 
forward ($g_2$) and umklapp ($g_3$) scattering. For repulsive interactions
$g_2,g_3\geq 0$. We do not consider backward scattering ($g_1$), since it
does not play any role in the mean field theory of the FISDW phases.

\subsection{Bare susceptibility}

The one-particle eigenstates of ${\cal H}_0$ were obtained 
in Ref.\ \onlinecite{ND95}  in the particular case where $t_\perp
(u)=-2t_b\cos (u)$. The extension to a general $t_\perp (u)$ is
straightforward, and we only quote the final result. The eigenstates and the
spectrum can be written as 
\begin{eqnarray}
\phi _{k_x,l}^\alpha ({\bf r}) &=& \frac{1}{\sqrt{bL_x}} e^{ik_xx} 
f_{l-m}^\alpha , \nonumber \\
 f_{l-m}^\alpha &=& \int _0^{2\pi } \frac{du}{2\pi }
 e^{-i(l-m)u+i\frac{\alpha }{\omega _c} T_\perp (u)}, \nonumber \\
\epsilon _{k_x,l,\sigma }^\alpha &=& v_F(\alpha k_x-k_F) + \alpha l \omega
 _c +\sigma \mu _BH, 
\label{phidef}
\end{eqnarray} 
where we have introduced 
\begin{equation}
T_\perp (u)= \int _0^u du' t_\perp (u') .
\end{equation} 
$L_x$ is the length of the system in the $x$ direction. $k_x$ is the
eigenvalue of the operator $-i\partial _x$ (which commutes with ${\cal
H}_0$). The wave function $\phi
_{k_x,l}^\alpha $ is localized around the $l$th chain with a spatial
extension in the $y$ direction of the order of $bt_b/\omega _c$ (assuming
$t_b\gg t_{2b},t_{3b}\cdots $). $bt_b/\omega _c$ also corresponds to the
amplitude of the semiclassical orbits in the transverse direction. The
localization of the wave functions can be interpreted as Bloch
oscillations of the electrons in the magnetic field, and within the same
picture the quantized spectrum $\epsilon ^\alpha _{k_x,l,\sigma }$ can be
seen as a Wannier-Stark ladder. \cite{ND95} The latter  provides a very
natural picture of the quantized nesting mechanism, which is at the origin of
the FISDWs  in quasi-1D conductors. \cite{rev,Heritier84,ND95,Montambaux86} 
Indeed, for two rods $l_1$ and $l_2$ of the Wannier-Stark ladder, 
the ``nesting condition'' $\epsilon ^-_{k_x,l_1,\uparrow }=-\epsilon
^+_{k_x+Q_x,l_2,\downarrow }$ is fulfilled if $Q_x=2k_F+(l_1-l_2)G$.
Therefore we expect the formation of a SDW at a quantized wave vector
$Q_x=2k_F+NG$ ($N$ integer) in presence of repulsive electron-electron
interactions.   

Now we introduce the bare transverse spin susceptibility in the Matsubara
formalism 
\begin{eqnarray}
&& \chi ^{(0)}_\alpha ({\bf r}-{\bf r}',\tau ) = \nonumber \\ &&
\langle T_\tau 
\psi _{\bar \alpha \uparrow }^\dagger ({\bf r},\tau ) 
\psi _{\alpha \downarrow } ({\bf r},\tau )
\psi _{\alpha \downarrow }^\dagger ({\bf r}',0) 
\psi _{\bar \alpha \uparrow } ({\bf r}',0) \rangle 
\end{eqnarray}
which is to be calculated with ${\cal H}_0$ only. Here $\tau $ is an
imaginary time. In Fourier space, we obtain
\begin{eqnarray}
\chi ^{(0)}_\alpha ({\bf q},\omega _\nu ) &=& \frac{1}{L_xL_y} \int d^2r d^2r'
\int _0^{1/T} d\tau \, 
e^{-i{\bf q}\cdot ({\bf r}-{\bf r}')+i\omega _\nu \tau } 
\nonumber \\ && \times 
\chi ^{(0)}_\alpha ({\bf r}-{\bf r}',\tau )
\nonumber \\
&=& - \frac{T}{L_xL_y} \sum _{\omega } \int d^2r d^2r'
e^{-i{\bf q}\cdot ({\bf r}-{\bf r}')}\nonumber \\ && \times 
G_{\bar \alpha \uparrow } ({\bf r}',{\bf r},\omega )
G_{\alpha \downarrow } ({\bf r},{\bf r}',\omega +\omega _\nu ), 
\end{eqnarray} 
where $L_xL_y$ is the area of the system and $q_x\sim \alpha 2k_F$. 
$G_{\alpha \sigma }$ is the one-particle Green's function. $\omega =\pi
T(2n+1)$ ($n$ integer) is a fermionic Matsubara frequency, and $\omega _\nu
=2\pi T\nu $ ($\nu $ integer) a bosonic Matsubara frequency. Using
\begin{equation}
G_{\alpha \sigma }({\bf r},{\bf r}',\omega )= \sum _{k_x,l} 
\frac{\phi ^\alpha _{k_x,l}({\bf r})\phi ^{\alpha *}_{k_x,l}({\bf
r}')}{i\omega -\epsilon ^\alpha _{k_x,l,\sigma }},
\label{GFun} 
\end{equation}
we obtain the well known result \cite{rev}
\begin{equation}
\chi ^{(0)}_\alpha ({\bf q},\omega _\nu ) = \sum _n I_n^2(q_y) 
\chi ^{1D}_\alpha (q_x-\alpha nG,\omega _\nu ),
\end{equation}
where $\chi ^{1D}_\alpha (q_x,\omega _\nu )$ is the susceptibility of a
one-dimensional system without interaction. In the static limit ($\omega
_\nu =0$),
\begin{eqnarray}
\chi ^{1D}_\alpha (q_x)&=& \frac{N(0)}{2} \Biggl \lbrack \ln \left ( 
\frac{2\gamma E_0}{\pi T} \right ) + \Psi \left ( \frac{1}{2} \right ) 
\nonumber \\ && 
- {\rm Re}\, \Psi \left ( \frac{1}{2} +\frac{v_F}{4i\pi T} (q_x-\alpha 2k_F)
\right ) \Biggr \rbrack , 
\label{chi1D}
\end{eqnarray}
where $N(0)=1/\pi v_Fb$ is the density of states per spin, $\Psi $ the
digamma function, and $\gamma \simeq 1.781$ the exponential of the Euler
constant. $E_0$ is an ultraviolet cutoff of the order of the
bandwidth. Since $\chi ^{1D}_+(2k_F)=(N(0)/2)\ln (2\gamma E_0/\pi
T)$, the bare susceptibility $\chi ^{(0)}$ has logarithmic divergences at
quantized values $Q_x^{(N)}=2k_F+NG$ ($N$ integer) of the longitudinal wave
vector.  
The coefficients $I_n(q_y)$ defined by 
\begin{eqnarray}
I_{l-l'}(q_y) &=&  e^{i(l+l')q_yb/2} \sum _m  e^{-imq_yb} 
f^-_{l-m} f^+_{l'-m} \nonumber \\
&=& \int _0^{2\pi } \frac{du}{2\pi } e^{i(l-l')u+\frac{i}{\omega _c}
\lbrack T_\perp (u+q_yb/2)+T_\perp (u-q_yb/2) \rbrack } \nonumber \\
\label{Idef}
\end{eqnarray} 
are well known in the QNM.\cite{rev} They
crucially depend on the detailed structure of the quasi-1D Fermi surface and
therefore determine the stability of the metallic phase with respect to the
formation of a SDW phase.

\subsection{RPA susceptibility and phase diagram}

\begin{figure}
\epsfxsize 15cm 
\epsffile[-50 280 700 540]{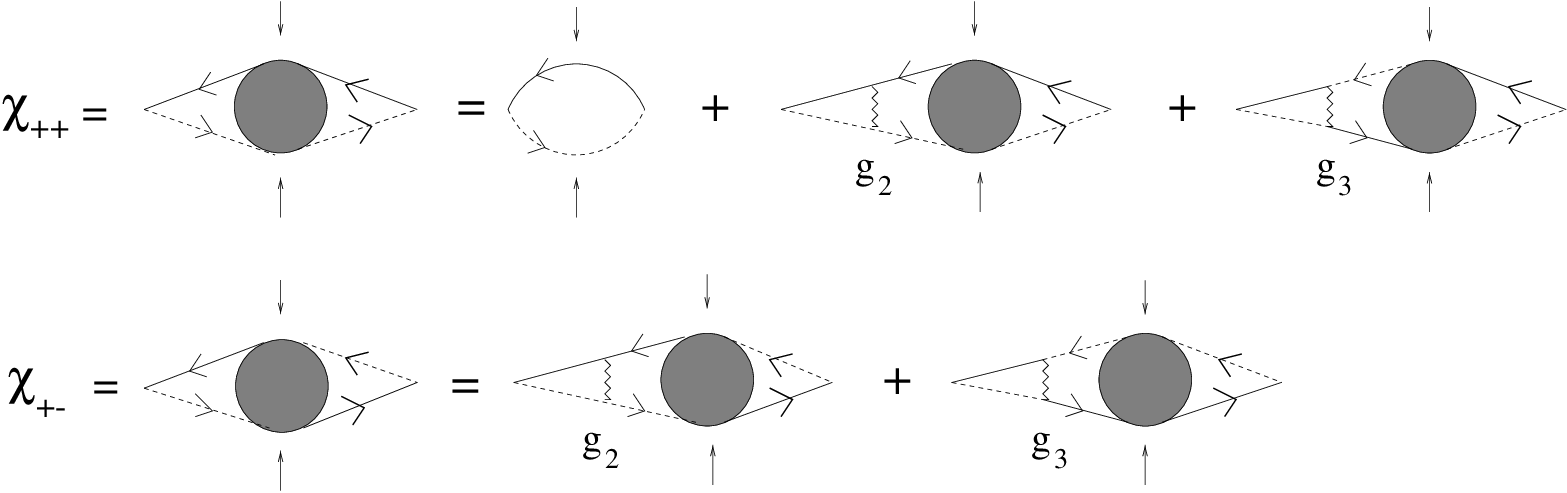}
\refstepcounter{figure} \label{Fig:chi}
{\small  FIG.\ \ref{Fig:chi}.
Diagrammatic representation of the integral equation for the
susceptibilities  $\chi _{++}$ and $\chi _{+-}$ in the RPA. 
The solid (dashed) lines represent electrons on the
right (left) sheet of the Fermi surface. The arrows indicate the spin
direction. The wavy lines correspond to forward ($g_2$) or
umklapp scattering ($g_3$). }  
\end{figure}

We now consider the total Hamiltonian as given by (\ref{ham1}), and
introduce the spin susceptibility 
\begin{eqnarray}
&& \chi _{\alpha \alpha '}({\bf r},{\bf r}',\tau ) = \nonumber \\ &&
\langle T_\tau 
\psi _{\bar \alpha \uparrow }^\dagger ({\bf r},\tau ) 
\psi _{\alpha \downarrow }({\bf r},\tau )
\psi _{\alpha '\downarrow }^\dagger ({\bf r}',0) 
\psi _{\bar \alpha '\uparrow } ({\bf r}',0) \rangle .
\end{eqnarray}
In the RPA, $\chi _{\alpha \alpha '}$ is determined by the integral equation
(see Fig.\ \ref{Fig:chi})
\begin{eqnarray}
\chi _{\alpha \alpha '}({\bf r},{\bf r}') &=& \delta _{\alpha ,\alpha '}
\chi ^{(0)}_\alpha ({\bf r}-{\bf r}') \nonumber \\ && 
+g_2 \int d^2 r_1 \chi ^{(0)}_\alpha ({\bf r}-{\bf r}_1) 
\chi _{\alpha \alpha '}({\bf r}_1,{\bf r}') \nonumber \\ && 
+g_3 \int d^2 r_1 \chi ^{(0)}_\alpha ({\bf r}-{\bf r}_1) e^{i\alpha 4k_Fx_1}  
\chi _{\bar \alpha \alpha '}({\bf r}_1,{\bf r}') \nonumber \\
\end{eqnarray}
where we now consider only the static limit ($\omega _\nu =0$). 
In Fourier space, this leads to
\begin{eqnarray}
\chi _{\alpha \alpha '}(q_x,q_x') &=& \delta _{\alpha ,\alpha '}
 \delta _{q_x,q_x'}  \chi ^{(0)}_\alpha (q_x) \nonumber \\ &&
+g_2 \chi ^{(0)}_\alpha (q_x) \chi _{\alpha \alpha '}(q_x,q_x') \nonumber \\ &&
+g_3 \chi ^{(0)}_\alpha (q_x) \chi _{\bar \alpha \alpha '}(q_x-\alpha
4k_F,q_x').
\end{eqnarray}
We have not written explicitly the dependence on $q_y$ since the
latter is a conserved quantity. Because $\chi ^{(0)}$ logarithmically
diverges at $Q_x^{(N)}=2k_F+NG$, we consider spin fluctuations only at these
wave vectors. Because of umklapp processes, fluctuations at $Q_x^{(N)}$ are
coupled with fluctuations at $Q_x^{(N)}-4k_F=-Q_x^{(\bar N)}$:
\begin{eqnarray}
\chi _{++}(Q_x^{(N)},Q_x^{(N)}) &=& \chi ^{(0)}_+(Q_x^{(N)}) \nonumber \\ &&
+g_2 \chi ^{(0)}_+(Q_x^{(N)})\chi _{++}(Q_x^{(N)},Q_x^{(N)})  \nonumber \\ &&
+g_3 \chi ^{(0)}_+(Q_x^{(N)}) \chi _{-+}(-Q_x^{(\bar N)},Q_x^{(N)}) ,
\nonumber \\
\chi _{-+}(-Q_x^{(\bar N)},Q_x^{(N)})  &=& g_2 \chi ^{(0)}_-(-Q_x^{(\bar N)}) 
\chi _{-+}(-Q_x^{(\bar N)},Q_x^{(N)}) \nonumber \\ && 
+g_3 \chi ^{(0)}_-(-Q_x^{(\bar N)}) \chi _{++}(Q_x^{(N)},Q_x^{(N)}) .
\nonumber \\
\end{eqnarray}
Using $\chi ^{(0)}_{\bar \alpha }(-q_x,q_y)=\chi ^{(0)}_\alpha (q_x,q_y)$,
we obtain 
\bleq 
\begin{eqnarray}
\chi _{++}({\bf Q}_N,{\bf Q}_N) &=& \frac{\chi ^{(0)}_+({\bf Q}_N)
\lbrack 1-g_2 \chi ^{(0)}_+({\bf Q}_{\bar N})\rbrack }
 { \lbrack 1-g_2\chi ^{(0)}_+({\bf Q}_N)\rbrack 
\lbrack 1-g_2\chi ^{(0)}_+({\bf Q}_{\bar N})\rbrack  
 -g_3^2  \chi ^{(0)}_+({\bf Q}_N) \chi ^{(0)}_+({\bf Q}_{\bar N}) }
 , \nonumber \\
\chi _{-+}(-Q_x^{(\bar N)},Q_y;{\bf Q}_N)  &=& \frac{g_3 \chi ^{(0)}_+({\bf
 Q}_{N}) \chi ^{(0)}_+({\bf Q}_{\bar N})} { \lbrack 1-g_2\chi ^{(0)}_+({\bf
Q}_N)\rbrack  
\lbrack 1-g_2\chi ^{(0)}_+({\bf Q}_{\bar N})\rbrack  
 -g_3^2  \chi ^{(0)}_+({\bf Q}_N) \chi ^{(0)}_+({\bf Q}_{\bar N}) } .
\label{chiRPA}
\end{eqnarray}
\eleq 
We have written explicitly the dependence on the transverse wave vector by
introducing 
\begin{equation} {\bf Q}_N=(Q_x^{(N)},Q_y)\,\, {\rm and} \,\, 
{\bf Q}_{\bar N}=(Q_x^{(\bar N)},-Q_y).
\label{QNdef}
\end{equation}
Note that in our notation the wave vector ${\bf Q}_{\bar N}\equiv {\bf Q}_{-N}$
has both the signs of $N$ and $Q_y$ reversed compared to ${\bf Q}_N$. This
happens because Umklapp scattering couples $(Q_x^{(N)},Q_y)$ to
$(-Q_x^{(\bar N)},Q_y)$, but the latter is equivalent to $(Q_x^{(\bar
N)},-Q_y)$. 

In presence of umklapp processes, the transition temperature $T_c^{(N)}$ is
determined by the modified Stoner criterion
\begin{eqnarray}
\lbrack 1-g_2\chi ^{(0)}_+({\bf Q}_N)\rbrack 
\lbrack 1-g_2\chi ^{(0)}_+({\bf Q}_{\bar N})\rbrack \nonumber \\ 
 -g_3^2  \chi ^{(0)}_+({\bf Q}_N) \chi ^{(0)}_+({\bf Q}_{\bar N}) = 0,
\label{stoner}
\end{eqnarray}
which is the condition for vanishing of denominators and divergence
of susceptibilities $\chi _{\alpha \alpha '}$ in (\ref{chiRPA}).
Two fluctuation modes diverge
simultaneously, which leads to the formation of two SDWs with wave vectors
${\bf Q}_N=(Q_x^{(N)},Q_y)$ and ${\bf Q}_{\bar N}=(Q_x^{(\bar N)},-Q_y)$
(see (\ref{chiRPA})). We label 
each FISDW phase by the integer $N$ such that the SDW with wave vector ${\bf
Q}_N$ has the largest amplitude. This is equivalent to defining $N$ by the
condition $\chi ^{(0)}_+({\bf Q}_N)>\chi ^{(0)}_+({\bf Q}_{\bar N})$ 
(section \ref{sec:polar}).

We will show in section \ref{sec:polar} that the SDWs can be either
sinusoidal or helicoidal, depending on the value of the angle $\vartheta _N$
($\in \rbrack -\pi /4,\pi /4\rbrack $)
defined by $\tan (2\vartheta _N)=2B/(A_N-A_{\bar N})$ and 
\begin{eqnarray}
A_{\pm N} &=&  \frac{1}{I_{\pm N}^2(Q_y)}
\left ( \frac{g_2}{g_2^2-g_3^2} - \chi ^{(0)}_+({\bf Q}_{\pm N}) \right ) ,
\nonumber \\ 
B &=& -\frac{g_3}{I_{N}(Q_y)I_{\bar N}(Q_y)(g_2^2-g_3^2)} .
\label{a12b}
\end{eqnarray}
For $\sin ^2(2\vartheta _N)<2/3$, the SDWs are sinusoidal and the Hall
effect is quantized: $\sigma _{xy}=-2Ne^2/h$. $N$ then corresponds to the
quantum number which is directly measured in transport experiments. 
For $\sin ^2(2\vartheta _N)>2/3$, the phase is helicoidal. The Hall effect
vanishes, but a kinetic magnetoelectric effect appears (see sections
\ref{sec:qhe} and \ref{sec:mee}). The phase $N=0$ is sinusoidal if $Q_y=\pi
/b$ and helicoidal if $Q_y\ne \pi /b$ (section \ref{sec:polar}). $\vartheta _N$
also determines the ratio $\vert \gamma \vert $ ($0<\vert \gamma \vert <1$)
of the amplitudes of the two SDWs: 
\begin{equation}
\gamma =\frac{I_N(Q_y)\tan (\vartheta _N)-rI_{\bar N}(Q_y)}{I_{\bar N}(Q_y)
-rI_N(Q_y)\tan (\vartheta _N)},
\end{equation}
where $r=g_3/g_2$. The figures of this section also show $\vert \tilde
\gamma \vert =|\tan (\vartheta _N)|$ which is
analogous to $\vert \gamma \vert $ but for the effective mean-field potential
acting on the electrons. ($\gamma $ and $\tilde \gamma $ are precisely
defined in section \ref{sec:polar}.) $\vert \tilde \gamma \vert $ and 
$\vert \gamma \vert $ increase with $|\vartheta _N|$. At the transition between
the sinusoidal and helicoidal phases $\vert \gamma \vert \sim \vert \tilde
\gamma \vert \simeq 0.518$.

In the absence of umklapp processes, Eq.\ (\ref{stoner}) yields the usual
Stoner criterion 
$1-g_2\chi ^{(0)}_+({\bf Q}_N)=0$ for the formation of a SDW at wave
vector ${\bf Q}_N=(Q_x^{(N)},Q_y)$. The quantized longitudinal wave vector
$Q_x^{(N)}=2k_F+NG$ and the transverse wave vector $Q_y$ are chosen to
maximize the transition temperature $T_c^{(N)}$ at a given magnetic field. 
Except when $N=0$, $Q_y$ is incommensurate: $Q_y\ne \pi/b$. The SDW is
sinusoidal ($\vartheta _N=0$ for $g_3=0$), and the quantum Hall conductivity in
the FISDW is determined by the integer $N$: $\sigma _{xy}=-2Ne^2/h$. As the
magnetic field increases, the value of 
$N$ changes, which leads to a cascade of FISDW phases separated by first
order transitions. In the simplest
version of the QNM, where $t_{3b}=t_{4b}=\cdots =0$, the phases of the
cascade are labeled by $N=\cdots 5,4,3,2,1,0$ as the field increases. 
The integer $N$ is positive provided $t_{2b}>0$, i.e.,
sgn($N$)=sgn($t_{2b}$). (For $t_{2b}<0$, one would obtain a similar 
sequence but with negative values of $N$.)

\begin{figure}
\epsfxsize 7.cm 
\epsffile[80 420 365 720]{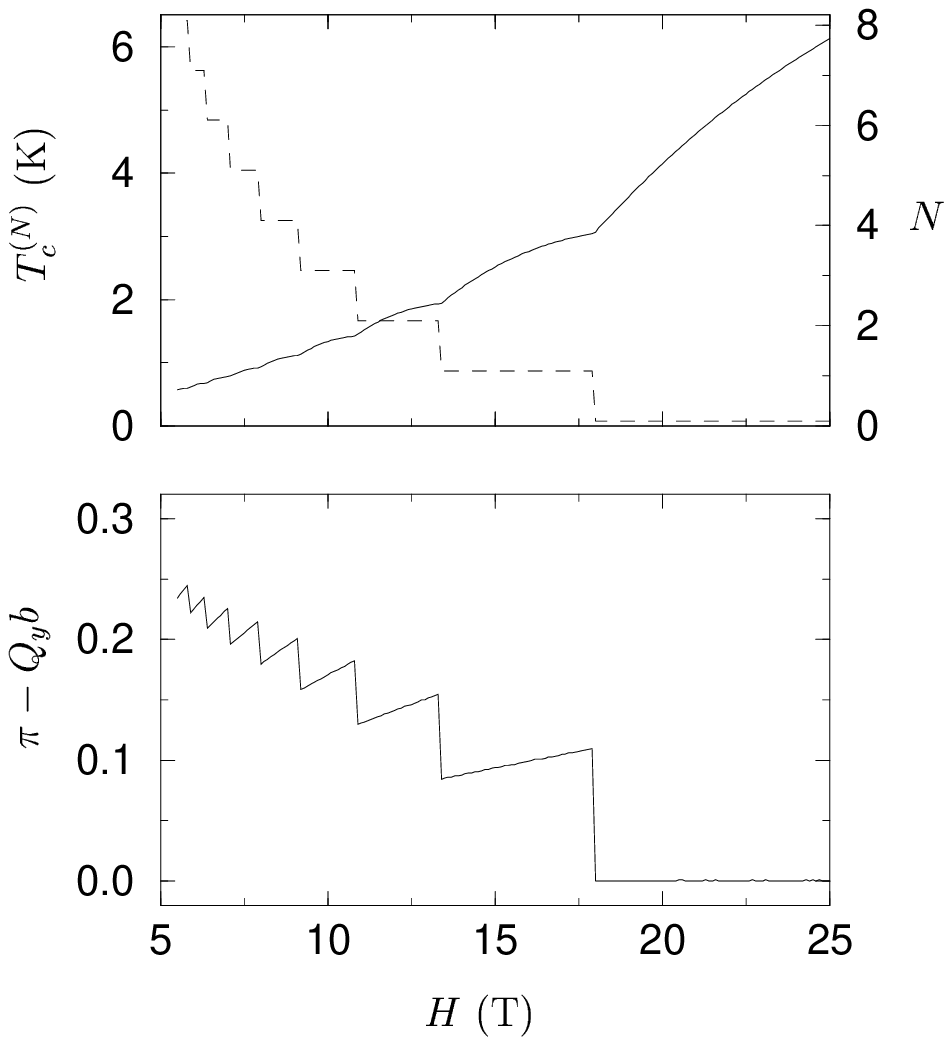}
\refstepcounter{figure} \label{Fig:r0}
{\small FIG.\ \ref{Fig:r0}.
 Phase diagram for $r=0$ ($\tilde g_2\simeq 0.38$ and $\tilde
g_3=0$) and $t_{4b}=0.75$ K. Top picture: Transition temperature $T_c^{(N)}$
(solid line). The dashed line gives the value of the integer 
$N$.  The  QHE is determined by $\sigma _{xy}=-2Ne^2/h$.
Bottom picture: Transverse wave vector $Q_y$ maximizing the
transition temperature  $T_c^{(N)}$. }
\end{figure}

\begin{figure}
\epsfxsize 7.cm 
\epsffile[80 290 365 745]{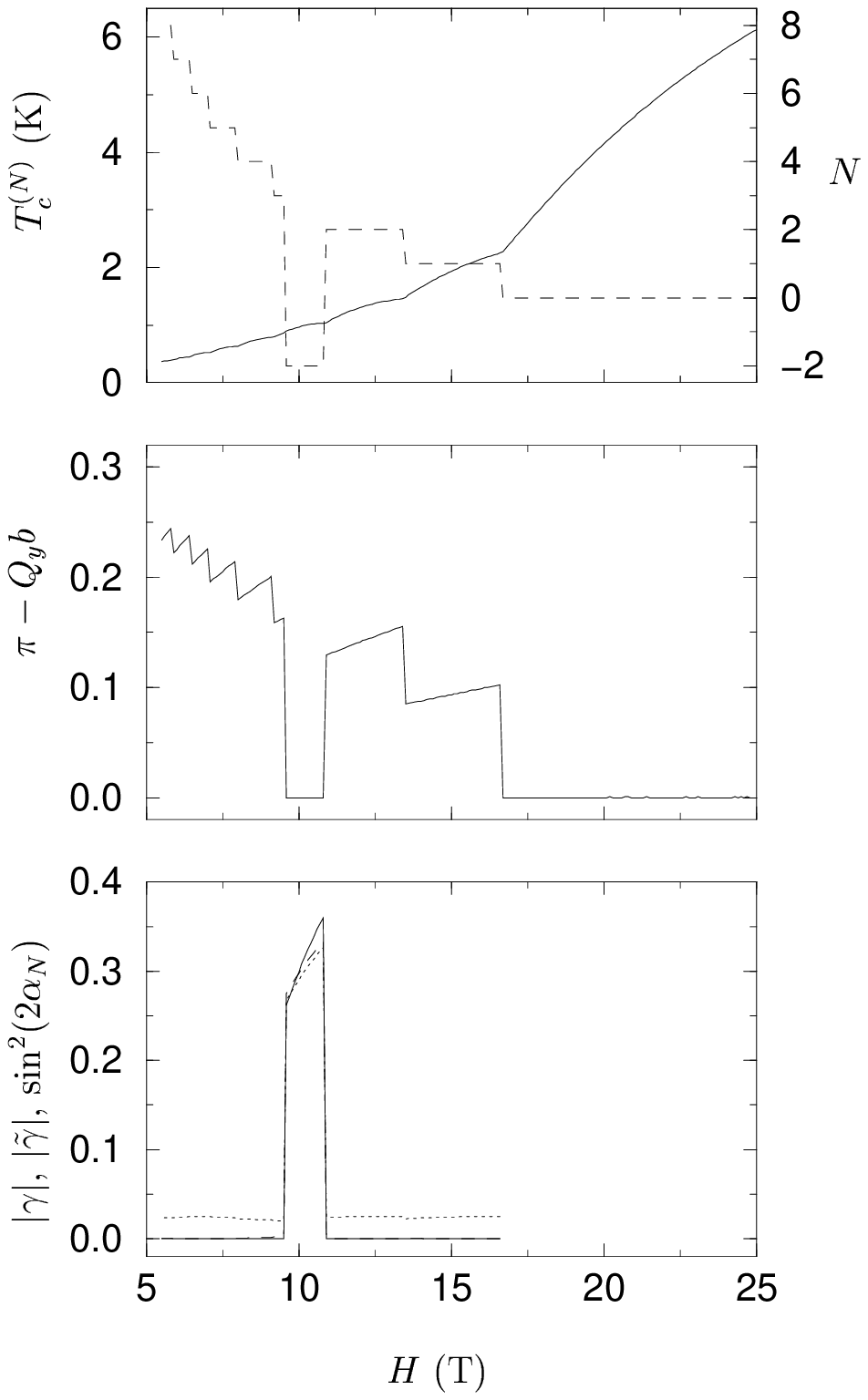}
\refstepcounter{figure} \label{Fig:r0v025}
{\small FIG. \ref{Fig:r0v025}.  
Phase diagram for $r=0.025$ ($\tilde g_2\simeq 0.37$ and $\tilde
g_3\simeq 0.01$) and $t_{4b}=0.75$ K. The bottom picture shows $\sin
^2(2\vartheta _N)$ (solid line), $|\tilde \gamma |$ (dashed line), and 
$|\gamma |$ (dotted line). $\sin ^2(2\vartheta _N)$ determines the polarization
of the SDWs and $|\tilde \gamma |$,$|\gamma |$ give the ratio of the
amplitudes of the two SDWs (see text for a precise definition). 
(These quantities are not shown in the last phase
of the cascade ($N=0$, $Q_y=\pi /b$) which contains
a single circularly polarized SDW.) A negative phase ($N=-2$) appears in
the cascade. The two SDWs have comparable amplitudes in that phase: 
$|\tilde \gamma |,|\gamma |\sim 0.3$. 
All the phases are sinusoidal (since $\sin ^2(2\vartheta _N)<2/3$)
and the Hall effect is quantized: $\sigma _{xy}=-2Ne^2/h$.   } 
\end{figure}

\begin{figure}
\epsfxsize 7.cm 
\epsffile[80 290 365 745]{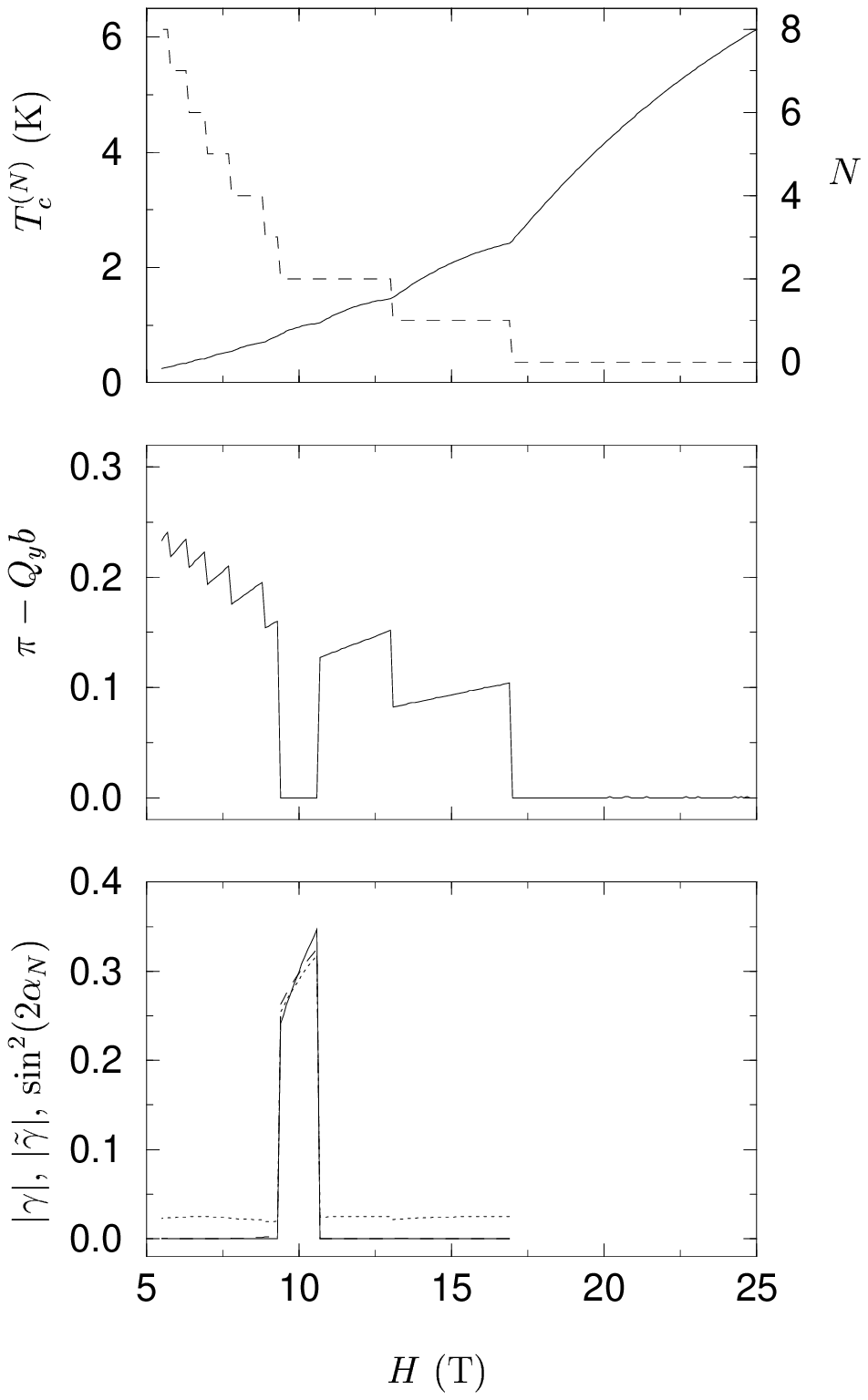}
\refstepcounter{figure} \label{Fig:r0v025bis}
{\small FIG. \ref{Fig:r0v025bis}. 
Phase diagram for $r=0.025$ ($\tilde g_2\simeq 0.37$ and $\tilde
g_3\simeq 0.01$) and $t_{4b}=-0.75$ K. All the phases are positive, but the
phase $N=2$ has split into two subphases, one of which is commensurate in
the transverse direction ($Q_y=\pi /b$). }
\end{figure}

\begin{figure}
\epsfxsize 7.cm 
\epsffile[80 290 365 745]{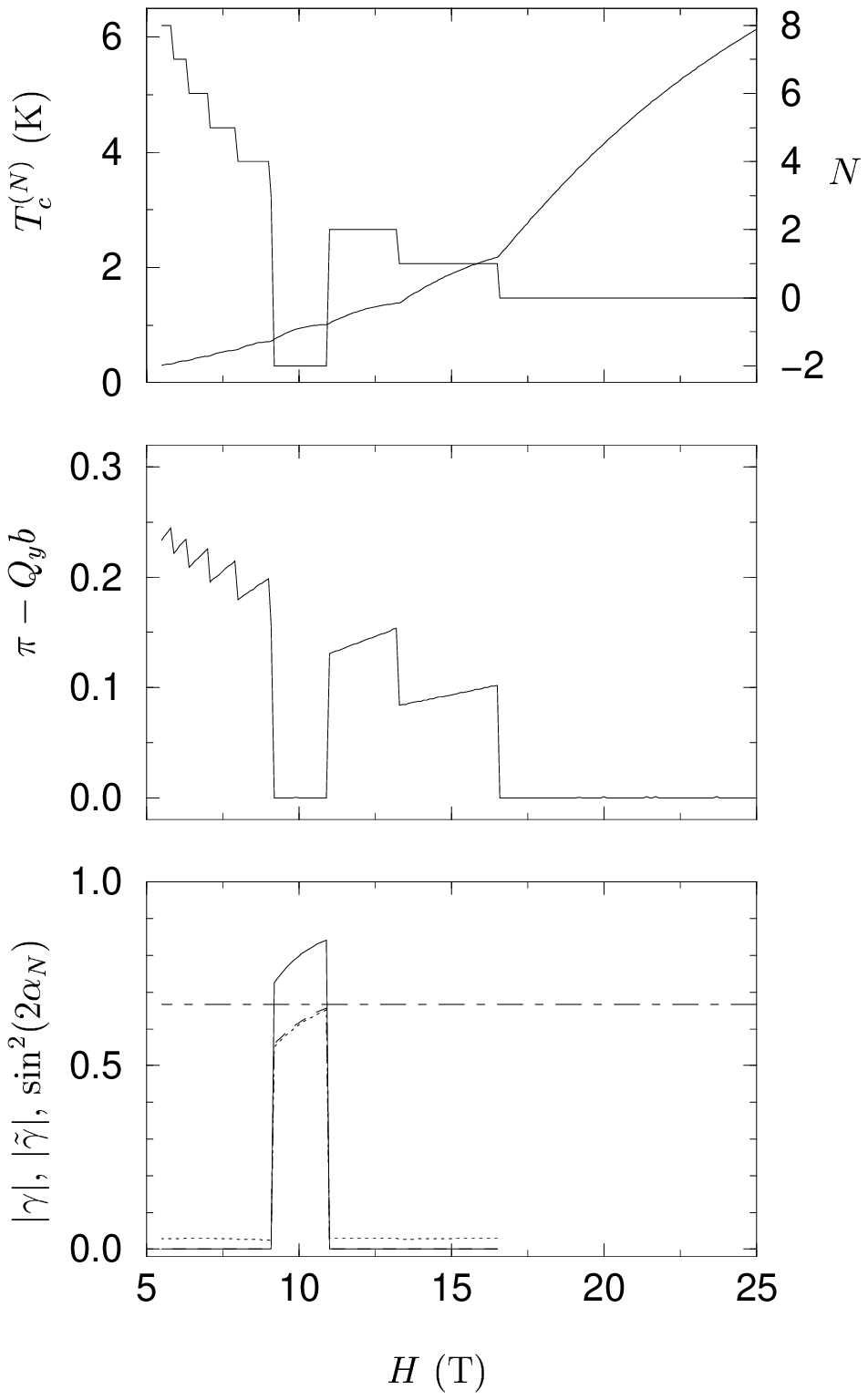}
\refstepcounter{figure} \label{Fig:t4b0v3}
{\small FIG. \ref{Fig:t4b0v3}. 
Phase diagram for $r=0.03$ and $t_{4b}=0.3$ K (the horizontal dot-dashed
line corresponds to 2/3). When $t_{4b}$ is
reduced from 0.75 to 0.3 K, the negative phase $N=-2$ becomes helicoidal
($\sin ^2(2\vartheta _{-2})>2/3$). The QHE vanishes in that phase, but a
magnetoelectric effect appears. }
\end{figure}

\begin{figure}
\epsfxsize 7.cm 
\epsffile[80 290 365 745]{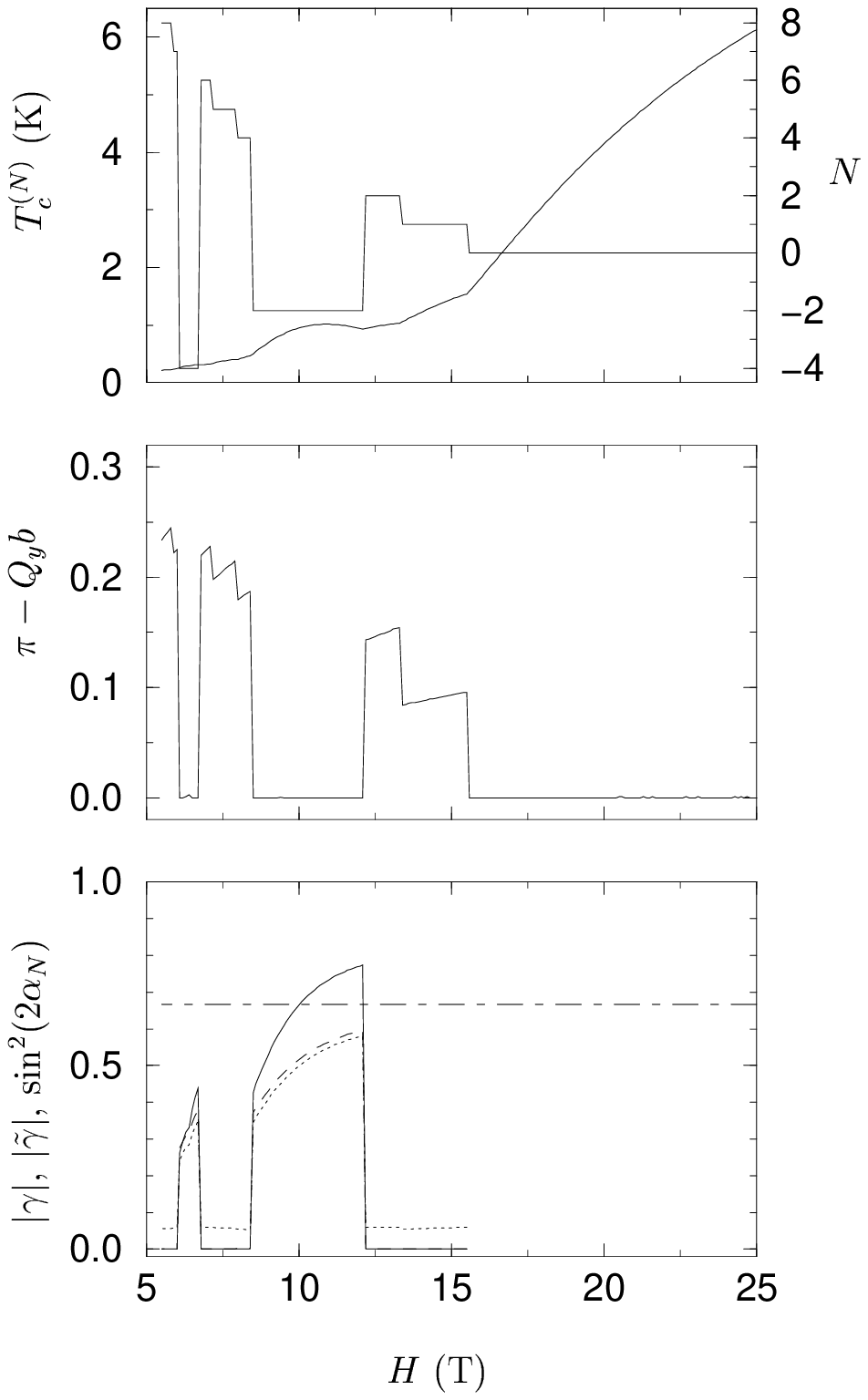}
\refstepcounter{figure} \label{Fig:r0v06}
{\small FIG. \ref{Fig:r0v06}. 
Phase diagram for $r=0.06$ and $t_{4b}=0.75$ K. Two negative
phases, $N=-2$ and $N=-4$, can be observed. The phase $N=-2$ has
split into two subphases: one is helicoidal ($\sin ^2(2\vartheta _{-2})>2/3$),
one is sinusoidal ($\sin ^2(2\vartheta _{-2})<2/3$). }  
\end{figure}

\begin{figure}
\epsfxsize 7.cm 
\epsffile[80 290 365 745]{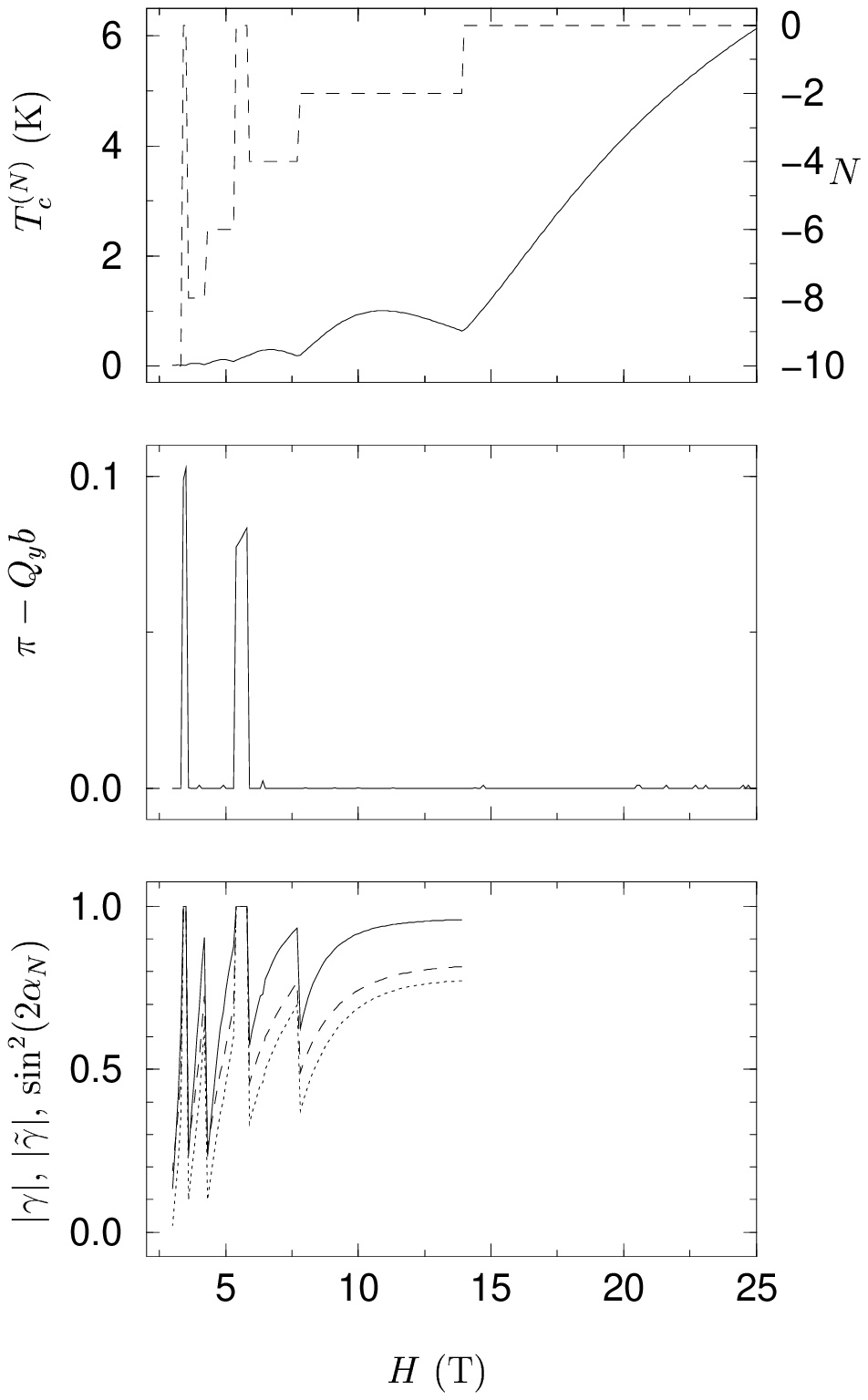}
\refstepcounter{figure} \label{Fig:r0v2}
{\small FIG. \ref{Fig:r0v2}. 
Phase diagram for $r=0.2$ ($\tilde g_2\simeq 0.32$ and $\tilde
g_3\simeq 0.06$) and $t_{4b}=0.75$ K. Only phases with negative even $N$
survive when umklapp scattering is strong enough. Some reentrances of the
phase $N=0$ (with $Q_y\ne \pi /b$)  also appear within the cascade. }
\end{figure}

We study the phase diagram in presence of umklapp scattering numerically
(see Figs.\ \ref{Fig:r0}--\ref{Fig:r0v2}).   
The calculations are done for $t_b=300$ K, $t_{2b}=20$ K, $t_{3b}=0$ K,
and $E_0=2000 $ K. $\tilde g_2+
\tilde g_3=2/\ln (2\gamma E_0/\pi T_c^\infty )$ is held fixed where
$T_c^\infty =12$ K  is the transition temperature for an infinite magnetic
field, and $\tilde g_i=N(0)g_i$ ($i=2,3$) are dimensionless coupling
constants. The figures are obtained for different values of $t_{4b}$ and
the ratio $r=\tilde g_3/\tilde g_2$. The parameters are such that we are in
the weak coupling limit: $\tilde g_3<\tilde g_2\lesssim 0.4$. In Bechgaard
salts, $t_{4b}$ is expected to be a very small energy scale, of order 1 K or
even less. \cite{Zanchi96} 

For $r=0$ and $t_{4b}=0.75$ K, we obtain the sequence $N=\cdots ,4,3,2,1,0$
in agreement with what has been found for $t_{4b}=0$ (i.e., a small value of
$t_{4b}$ does not change the phase diagram when $r=0$).\cite{rev}
The transverse wave vector $Q_y$ varies approximately
linearly with the field within each phase, and is incommensurate ($Q_y\ne
\pi /b$) except in the phase $N=0$ (Fig.\ \ref{Fig:r0}). 

A very small $g_3$ does not change the phase diagram qualitatively
compared to the case $g_3=0$. Now the main SDW at the wave vector
${\bf Q}_N$ coexists with a weak SDW at the wave vector ${\bf
Q}_{\bar N}$. In general, the value of $Q_y$ that maximizes $\chi ^{(0)}({\bf
Q}_N)$ does not maximize $\chi ^{(0)}({\bf Q}_{\bar N})$, so $\chi ^{(0)}({\bf 
  Q}_{\bar N})\ll\chi ^{(0)}({\bf Q}_N)$. As a result, the SDW amplitude at the
wave vector ${\bf Q}_{\bar N}$ is very small, and the polarizations of the
SDWs are linear. The values of $N$ follow the usual positive
sequence $N=\cdots ,5,4,3,2,1,0$ as the magnetic field increases.

A larger value of $g_3$ increases the coupling between the two SDWs. This
leads to a strong decrease of the critical temperature or even the 
disappearance of the SDWs. However, for even $N$, there exists a
critical value of $g_3$ above which the system prefers to choose the
transversely commensurate wave vector $Q_y=\pi/b$ for both SDWs. The
reason is that, for even $N$ (as opposed to odd $N$), $Q_y=\pi/b$
corresponds to a local maximum of the susceptibilities {\it and}
$\chi ^{(0)}(Q_x^{(N)},\pi/b)\simeq \chi ^{(0)}(Q_x^{(\bar N)},\pi/b)$.
The two susceptibilities are strictly equal at $t_{4b}=0$, but when
$t_{4b}>0$, $\chi ^{(0)}(Q_x^{(/bar N)},\pi/b)>\chi ^{(0)}(Q_x^{(N)},\pi/b)$
(this result also 
holds for $t_{3b}\neq0$ since $\chi ^{(0)}(Q_x^{(N)},\pi/b)$ is
independent of $t_{3b}$).\cite{Zanchi96} This yields a negative
Hall plateau, provided the SDWs are sinusoidal. Thus, for
$r=0.025$ ($\tilde g_2\simeq 0.37$ and $\tilde
g_3\simeq 0.01$) and $t_{4b}=0.75$ K,
we find the sequence $N=\cdots 5,4,3,-2,2,1,0$ (Fig.\ 
\ref{Fig:r0v025}).  A negative
commensurate phase with $N=-2$ and $Q_y=\pi/b$ appears in the cascade.
All the phases are sinusoidal, so the Hall effect is quantized
($\sigma_{xy}=-2Ne^2/h$). For $r=0.025$ and $t_{4b}=-0.75$ K, we
obtain only positive integers: $N=\cdots 5,4,3,2,1,0$ (Fig.\
\ref{Fig:r0v025bis}). However, the phase $N=2$ has split into two subphases:
for large enough field, the transverse wave 
vector $Q_y$  differs from $\pi /b$ and varies linearly with the
field; for a weaker field, the phase is commensurate
($Q_y=\pi /b$). All the phases are sinusoidal ($\sin ^2(2\vartheta _N)<2/3$) so
that the Hall effect is quantized. 

Fig.\ \ref{Fig:r0v025} shows that both
SDWs have comparable amplitudes in the negative phases (and more generally
in the phases where $Q_y=\pi /b$ (Fig.\ \ref{Fig:r0v025bis})): 
$|\gamma |,|\tilde \gamma |\simeq 0.3$. This results
from the property $\chi ^{(0)}(Q_x^{(N)},\pi/b)\simeq \chi ^{(0)}(Q_x^{(\bar
N)},\pi/b)$. On the contrary, the amplitude of the SDW with wave vector
${\bf Q}_{\bar N}$ remains very small in the positive phases. Note that it
is $\tilde \gamma $ which is actually vanishingly small, $|\gamma |$ being
of the order of $r$ (see section \ref{subsec:polarA} for a further
discussion).  

The strength of umklapp scattering is very sensitive to pressure. Indeed,
hydrostatic pressure reduces the dimerization gap and diminishes $g_3$. 
Therefore, we conclude
that sign reversals of the QHE can be induced by varying pressure. In our
simplified model, this effect requires $t_{4b}>0$. Our results
provide a new explanation of the sign reversals of the QHE which
have long been observed in quasi-1D organic conductors. 
\cite{Ribault85,Piveteau86,Hannahs89,Cooper89,Balicas95} 
In particular, Balicas {\it et
al.} have recently shown unambiguously the existence of the phase $N=-2$ 
in (TMTSF)$_2$PF$_6$ at a
pressure of 8.3 kbar by observing a sign reversal of the QHE with a well
defined Hall plateau corresponding to $N=-2$. \cite{Balicas95}
(These results required a conditioning procedure
in which current pulses depin the FISDW from lattice defects and tend to
reduce hysteresis.) The observed FISDW cascade corresponds to
$N=\cdots 4,3,-2,2,1,0$. When 
the pressure is increased to 9 kbar (which decreases the
umklapp scattering strength), the phase $N=-2$ disappears and the usual
sequence $N=...4,3,2,1,0$ is obtained.

If the value of $t_{4b}$ is reduced, the
phase $N=-2$ becomes helicoidal. This is shown in Fig.\ \ref{Fig:t4b0v3}
obtained for $r=0.03$ and $t_{4b}=0.3$ K. (When $t_{4b}=0.3$ K, the phase
$N=-2$ appears for stronger umklapp scattering. This is the reason why we
show the phase diagram for $r=0.03$ and not $r=0.025$.) In the
helicoidal phase, $|\gamma |\sim |\tilde \gamma |\gtrsim 0.5$. For $t_{4b}=0$,
there is a degeneracy between $N$ and $-N$ at $Q_y=\pi /b$:  $\chi
^{(0)}_+ (Q_x^{(N)},\pi /b)=\chi ^{(0)}_+(Q_x^{(\bar N)},\pi /b)$, $I_N(\pi
/b)=I_{\bar N}(\pi  
/b)$ and $A_N=A_{\bar N}$. This yields $\vartheta _N=\pi /4$ and the waves are
helicoidal. A finite $t_{4b}$ lifts this degeneracy, so that $A_N\neq
A_{\bar N}$ 
and in turn $\sin ^2 (2\vartheta _N)<1$. Thus, the stability of the helicoidal
phases is strongly related to the degeneracy between $N$ and $-N$ occuring
for $Q_y=\pi /b$. In our model, this degeneracy is entirely controlled
by $t_{4b}$. It is not affected by $t_{3b}$, which is the reason why we have
chosen $t_{3b}=0$ in the numerical calculations. 

If, on the other hand, the value of $t_{4b}$ is increased, the ratio of the
amplitudes of the two SDWs decreases. For instance, for $r=0.025$ and
$t_{4b}=1.5$ K, we find $|\gamma |,|\tilde \gamma |\sim 0.15$. Thus, for
large $t_{4b}$ (i.e., $t_{4b}\gtrsim 1.5$ K), the amplitude of the SDW with
wave vector ${\bf Q}_{\bar N}$ becomes very small. There is then no real
difference between a positive and a negative phase (except for the sign of the
QHE) insofar as both contain a main SDW, which coexists with another SDW with
a very small amplitude. 

If $r$ is increased to 0.06 (with $t_{4b}=0.75$ K), a second negative phase
($N=-4$) appears, and the cascade becomes $N=\cdots 8,7,-4,6,5,4,-2,2,1,0$
(Fig.\ \ref{Fig:r0v06}). Note that $N=-2$ and
$N=-4$ are the two negative phases  most easily  observed in
experiments.\cite{Cooper89,Zanchi96} The phase $N=-2$ has
split into two subphases: one is helicoidal ($\sin ^2(2\vartheta _2)>2/3$),
one is sinusoidal ($\sin ^2(2\vartheta _2)<2/3$). Thus, increasing the
strength of umklapp processes makes the negative phase $N=-2$ helicoidal. 

In order to observe the helicoidal phase
experimentally, it would be desirable to
stabilize the negative phase $N=-2$ at the lowest possible pressure (which
corresponds to the strongest $g_3$). In
(TMTSF)$_2$PF$_6$, the pressure has to be larger than 6 kbar, since below this
pressure the FISDW cascade disappears.\cite{rev} In the experiment
reported in Ref.\ \onlinecite{Balicas95}, where the phase $N=-2$ has been
observed at 8.3 kbar, the pressure could be reduced only by about 2
kbar. Nevertheless, because $g_3$ is very sensitive to pressure, such a
pressure reduction could induce a significant increase of the umklapp
scattering strength. (TMTSF)$_2$ReO$_4$, where sign reversals of the QHE
have been observed under pressure,\cite{Kang91} could also be a good
candidate for the observation of helicoidal phases. In (TMTSF)$_2$ClO$_4$,
sign reversals of the QHE have been observed at ambient pressure,
\cite{rev,Ribault85} so that it is not possible to increase $g_3$ by
decreasing pressure.  

When $r$ is further increased, only phases with negative even $N$ survive.
This leads to the sequence $N=\cdots -8,-6,-4,-2,0$ for $t_{4b}=0.75$ K apart
from some reentrances of the phase $N=0$ within the cascade to be discussed
below. (For  $t_{4b}=-0.75$ K, we obtain the sequence $N=\cdots 8,6,4,2,0$.)
Moreover, all the phases $N\ne 0$ are commensurate ($Q_y=\pi /b$). 
This is shown in Fig.\ \ref{Fig:r0v2} obtained for $r=0.2$, i.e., $\tilde
g_2\simeq 0.32$ and $\tilde g_3\simeq 0.06$.

The phase $N=0$ is somehow special since $\chi
^{(0)}_+({\bf Q}_N)=\chi ^{(0)}_+ ({\bf Q}_{\bar N})$ in that case. 
As a result, the transition temperature
$T_c^{(0)}$ is determined by $1-(g_2+g_3)\chi ^{(0)}_+({\bf Q}_0)=0$ and
does not 
depend on the ratio $r=g_3/g_2$ when $g_2+g_3$ is held fixed. This should be
contrasted with the transition temperature $T_c^{(N)}$ ($N\neq 0$) which
decreases with $r$ (except when $Q_y=\pi /b$ and  $\chi
^{(0)}_+(Q_x^{(N)},\pi /b)=\chi ^{(0)}_+(Q_x^{(\bar N)},\pi /b)$). 
This explains why, when 
$r$ increases, some reentrances of the phase $N=0$ are observed within the
cascade. Notice that the latter phases are not commensurate ($Q_y\neq \pi
/b$) contrary to the last phase ($N=0$, $Q_y=\pi /b$) of the cascade. 
The reentrant phases $N=0$ are always helicoidal
(independently of the structure of the Fermi surface), but the last phase
$N=0$, $Q_y=\pi /b$ of the cascade is sinusoidal (section \ref{sec:polar}).

\subsection{Effect of 1D fluctuations}

Our numerical results show that a very small value of the umklapp scattering
amplitude is sufficient to explain the phase diagram of Bechgaard
salts. In fact, the order of magnitude of $g_3$ which is required to
stabilize negative phases strongly depends on the choice of the ultraviolet
cutoff $E_0$. 

Within a mean-field picture, $E_0$ is a large energy, of the order of the
electron bandwidth. It has been argued that because of 1D fluctuations, the
appropriate cutoff to be used in the QNM is not the bandwidth but the
dimensional crossover temperature $T_{x^1}\ll E_0$.\cite{Hubert93} 
(Above $T_{x^1}$, the
behavior of the system is essentially 1D, so that the interference between
particle-particle and particle-hole channels invalidate the mean-field (or
ladder) approximation.) A stronger  value of $g_3$ is then required to
stabilize negative phases. This is in agreement with the suggestion of
Behnia {\it et al.} that the effective low-temperature value of $g_3$ in
a magnetic field is significantly enhanced by 1D high energy
scales. \cite{Behnia95} This point of view is supported by NMR measurements
and a large magneto-resistance, which shows an activated behavior becoming
more and more pronounced as the field increases.\cite{Behnia95}

\section{Order parameters: Helicoidal vs sinusoidal waves}
\label{sec:op}

The divergence of the susceptibilities $\chi _{\alpha \alpha '}$ (section 
\ref{sec:imp}) indicates that the FISDW phases are characterized by the order
parameters 
\begin{equation}
\Delta _{\alpha \sigma }({\bf r}) = \langle \psi _{\alpha \sigma }^\dagger
({\bf r}) \psi _{\bar \alpha \bar \sigma } ({\bf r}) \rangle
 = \sum _{p=\pm } \Delta _{\alpha \sigma }^{(pN)}e^{-i\alpha {\bf Q}_{pN} 
\cdot {\bf r}} .
\label{poR}
\end{equation} 
The two wave vectors ${\bf Q}_{pN}$, $p=\pm 1$, are given by Eq.\
(\ref{QNdef}).  $Q_x^{(pN)}=2k_F+pNG$ and $Q_y^{(\bar N)}=-Q_y^{(N)}$.
The complex numbers $\Delta _{\alpha \sigma }^{(pN)}$ are the order parameters
of the FISDW phase. $\Delta _{\alpha \sigma } 
({\bf r})=\Delta _{\bar \alpha \bar \sigma }^*({\bf r})$
implies  $\Delta _{\alpha \sigma }^{(pN)}= 
\Delta _{\bar \alpha \bar \sigma }^{(pN)*}$. 
Among the eight order parameters $\Delta _{\alpha \sigma }^{(pN)}$, only 
four are therefore independent and sufficient to characterize the SDW phase. 
Note that for $N=0$, one should distinguish in general between the phases
$N$ and $\bar N$. In this case, there are two SDWs with wave vectors
$(2k_F,Q_y^{(0)})$ and $(2k_F,Q_y^{(\bar 0)})$ with $Q_y^{(\bar
0)}=-Q_y^{(0)}$. For 
$Q_y^{(0)}\neq 0,\pi /b$, the two SDWs are different so that both 
$\Delta _{\alpha \sigma }^{(0)} $ and  $\Delta _{\alpha \sigma }^{(\bar 0)}$
are needed. When $Q_y^{(0)}=0,\pi /b$, the two SDWs are identical, and only
one order parameter (for instance $\Delta _{\alpha \sigma }^{(0)} $) is
sufficient.  

Now we discuss how the polarization of the wave affects the order parameters
$\Delta _{\alpha \sigma }^{(pN)}$. For simplicity, we consider only one wave
vector ${\bf Q}$ and denote the four different order parameters by $\Delta
_{\alpha \sigma }$ (among which only two are independent since $\Delta
_{\bar \alpha 
\bar \sigma }=\Delta _{\alpha \sigma }^*$). For a SDW polarized
perpendicularly to the magnetic field, the expectation value of the
spin density operator ${\bf S}({\bf r})$ can be written as
\begin{eqnarray}
\langle S_x({\bf r})\rangle &=& \sum _{\alpha ,\sigma ,\sigma '} 
\langle \psi ^\dagger _{\alpha \sigma }({\bf r}) 
\bbox{\tau }^{(x)}_{\sigma ,\sigma '} 
\psi _{\bar \alpha \sigma '}({\bf r})\rangle \nonumber \\ &=&
m_x \cos ({\bf Q}\cdot {\bf r}+\theta _1 ),
\nonumber \\ 
\langle S_y({\bf r})\rangle &=& \sum _{\alpha ,\sigma ,\sigma '} 
\langle \psi ^\dagger _{\alpha \sigma }({\bf r}) 
\bbox{\tau }^{(y)}_{\sigma ,\sigma '} 
\psi _{\bar \alpha \sigma '}({\bf r})\rangle \nonumber \\ &=&
m_y \cos ({\bf Q}\cdot {\bf r}+\theta _2 ) ,
\label{magne}
\end{eqnarray}
where $\bbox{\tau }^{(x)}$ and $\bbox{\tau }^{(y)}$ are Pauli matrices. 
$\theta _1=\theta _2$ corresponds to a sinusoidal wave, while $\theta
_1=\theta _2 \pm \pi /2$ and $m_x=m_y$ corresponds to a helicoidal wave. 
$\langle {\bf S}({\bf r})\rangle $ can be expressed in terms of the
order parameters (\ref{poR}): 
\begin{eqnarray} 
\langle S^+({\bf r})\rangle &=& \langle S_x({\bf r})\rangle + 
i\langle S_y({\bf r})\rangle \nonumber \\ 
&=& \sum _\alpha \langle \psi _{\alpha \uparrow }^\dagger ({\bf r}) 
\psi _{\bar \alpha \downarrow } ({\bf r}) \rangle 
=\sum _\alpha \Delta _{\alpha \uparrow } ({\bf
r}) .
\end{eqnarray}
Comparing this expression with Eqs.\ (\ref{magne}), we find
\begin{eqnarray}
\Delta _{+\uparrow } &=& \frac{1}{2} \left ( m_x e^{-i\theta _1} +im_y
e^{-i\theta _2} \right ), \nonumber \\
\Delta _{-\uparrow } &=& \frac{1}{2} \left ( m_x e^{i\theta _1} +im_y
e^{i\theta _2} \right ).
\label{DvsS}
\end{eqnarray}

For a sinusoidal wave $\vert \Delta _{+\uparrow }\vert =\vert \Delta
_{-\uparrow }\vert $. For a helicoidal wave $\Delta _{+\uparrow }\neq 0$
and $\Delta _{-\uparrow }=0$ (or the symmetric solution  $\Delta _{-\uparrow
}\neq 0$ and $\Delta _{+\uparrow }=0$). 
The inverse is also true:  $\vert \Delta _{+\uparrow }\vert
=\vert \Delta _{-\uparrow }\vert $ implies that the wave is sinusoidal, while
 $\Delta _{+\uparrow }\neq 0$ and $\Delta _{-\uparrow }=0$ implies that the
wave is helicoidal. .

\section{Polarization of the phase $N$}
\label{sec:polar}

In this section we derive  the Ginzburg-Landau expansion of the free
energy as a function of the order parameters $\Delta _{\alpha \sigma
}^{(pN)}$. The minimum of the free energy determines the polarization
(linear or circular) of the SDWs. 

The mean-field (or Hartree-Fock) Hamiltonian is given by 
\begin{eqnarray}
{\cal H}_{\rm MF} &=& {\cal H}_0 - \sum _{\alpha ,\sigma } \int d^2r 
\tilde \Delta _{\alpha \sigma } ({\bf r}) \psi _{\bar \alpha \bar
\sigma }^\dagger ({\bf r}) \psi _{\alpha \sigma } ({\bf r}) \nonumber \\ &&
+ \sum _\alpha  \int d^2r \tilde \Delta _{\alpha \uparrow }({\bf r})
\Delta _{\bar \alpha \downarrow }({\bf r})  ,
\label{ham2}
\end{eqnarray}
where we have introduced
\begin{equation}
\tilde \Delta _{\alpha \sigma } ({\bf r}) = g_2 \Delta _{\alpha \sigma } ({\bf
r}) +g_3 e^{-i\alpha 4k_Fx} \Delta _{\bar \alpha \sigma } ({\bf r}) .
\end{equation} 
$\Delta _{\alpha \sigma } ({\bf r})$ is given by (\ref{poR}). For $g_3\ne
0$, the mean-field potential $\tilde \Delta _{\alpha \sigma }({\bf r})$
acting on the electrons is a linear combination of the order parameters
$\Delta _{\alpha \sigma }({\bf r})$. 

Calculating the free energy (per surface unit) of the phase $N$ to the
fourth order in the 
order parameters, we obtain $F_N=F^{(2)}_N+F^{(4)}_N$ with 
\begin{eqnarray}
F^{(2)}_N&=&\sum _\alpha  \int \frac{d^2r}{L_xL_y} \tilde \Delta _{\alpha
\uparrow }  
({\bf r})\Delta _{\bar \alpha \downarrow } ({\bf r})
 \nonumber \\ && 
+\frac{T}{L_xL_y} \sum _{\alpha ,\omega } \int d^2 r_1 d^2r_2  \tilde \Delta
_{\alpha \uparrow } ({\bf r}_1) \tilde \Delta _{\bar \alpha \downarrow}({\bf
r}_2) \nonumber \\ && \times 
G_{\alpha \uparrow }({\bf r}_1,{\bf r}_2,\omega ) 
G_{\bar \alpha \downarrow }({\bf r}_2,{\bf r}_1,\omega ) ,
\nonumber \\ 
F^{(4)}_N&=& \frac{T}{2L_xL_y} \sum _{\alpha ,\omega } \int d^2r_1 d^2r_2
d^2r_3 d^2r_4  \tilde \Delta _{\alpha \uparrow } ({\bf r}_1)
\tilde \Delta _{\bar \alpha \downarrow }({\bf r}_2)
\nonumber \\ && \times 
\tilde \Delta _{\alpha \uparrow } ({\bf r}_3)
\tilde \Delta _{\bar \alpha \downarrow }({\bf r}_4)
G_{\alpha \uparrow }({\bf r}_1,{\bf r}_2,\omega ) 
G_{\bar \alpha \downarrow }({\bf r}_2,{\bf r}_3,\omega )
\nonumber \\ && \times 
G_{\alpha \uparrow }({\bf r}_3,{\bf r}_4,\omega ) 
G_{\bar \alpha \downarrow }({\bf r}_4,{\bf r}_1,\omega ) .
\end{eqnarray}
$G_{\alpha \sigma
}$ is the single-particle Green's function in the metallic phase (see
(\ref{GFun})). Introducing the electron-hole pairing amplitude (we use the
notation $w_i\equiv (k_{ix},l_i)$ for the indices of the functions 
$\phi _{k_{ix},l_i}^\alpha $ (Eq.\ (\ref{phidef})))  
\begin{eqnarray}
\tilde \Delta _{\alpha \sigma } (w_1,w_2) &=& \int d^2r \phi _{w_1} ^{\bar
\alpha *}({\bf r}) \phi _{w_2}^\alpha ({\bf r}) \tilde \Delta _{\alpha
\sigma } ({\bf r}) \nonumber
\\ &=& \tilde \Delta _{\bar \alpha \bar \sigma }^*(w_2,w_1) ,
\end{eqnarray}
we write the free energy as
\begin{eqnarray}
F_N^{(2)}&=& \sum _\alpha  \int \frac{d^2r}{L_xL_y}\tilde \Delta _{\alpha
\uparrow }  ({\bf r})\Delta _{\bar \alpha \downarrow } ({\bf r}) 
\nonumber \\ &&  
+ \frac{T}{L_xL_y} \sum _{\alpha ,\omega ,w_1,w_2} \frac{\tilde \Delta _{\alpha
\uparrow } (w_1,w_2) \tilde \Delta _{\bar \alpha \downarrow
}(w_2,w_1)}{(i\omega -\epsilon ^{\bar 
\alpha } _{w_1\downarrow })(i\omega -\epsilon ^\alpha _{w_2\uparrow })} ,
\end{eqnarray}
\bleq
\begin{equation}
F_N^{(4)}= \frac{T}{2L_xL_y} \sum _{\alpha ,\omega }\sum _{w_1,w_2,w_3,w_4}
\frac{\tilde \Delta _{\alpha \uparrow } (w_1,w_3)\tilde \Delta _{\bar \alpha
\downarrow }(w_3,w_4) \tilde  
\Delta _{\alpha \uparrow } (w_4,w_2)\tilde  \Delta _{\bar \alpha \downarrow
}(w_2,w_1)} 
{ (i\omega -\epsilon ^{\bar \alpha }_{w_1\downarrow })
(i\omega -\epsilon ^\alpha _{w_3\uparrow })
(i\omega -\epsilon ^{\bar \alpha }_{w_4\downarrow })
(i\omega -\epsilon ^\alpha _{w_2\uparrow })} . 
\end{equation}
\eleq
The pairing amplitudes are given by (appendix \ref{app:pa}) 
\begin{eqnarray}
\tilde \Delta _{\alpha \sigma } (w_1,w_2) &=& \sum _{p=\pm } \delta
_{k_{2x},k_{1x}+\alpha Q_x^{(pN)}} (g_2 \Delta _{\alpha \sigma }^{(pN)}
+g_3 \Delta _{\bar \alpha \sigma }^{(\bar pN)}) 
\nonumber \\ && \times 
e^{-i\alpha Q_y^{(pN)}b(l_1+l_2)/2} I_{\alpha (l_1-l_2)} (Q_y^{(N)}) .
\label{pa}
\end{eqnarray}
This leads to
\begin{eqnarray}
F^{(2)}_N &=& \sum _{p,\alpha } \Bigl \lbrack A_{pN} 
\vert \tilde \Delta _{\alpha \uparrow }^{(pN)} \vert ^2 
+B \tilde \Delta _{\alpha \uparrow }^{(pN)} 
\tilde \Delta _{\bar \alpha \uparrow }^{(\bar pN)*}
\Bigr \rbrack +\delta F_0^{(2)} , \nonumber \\ 
\delta F_0^{(2)} &=& 
\delta _{N,0} \Bigl ( \sum _n \delta _{Q_y^{(0)}b,n\pi } \Bigr )
\sum _\alpha 
\Bigl \lbrack A_0
( \tilde \Delta _{\alpha \uparrow }^{(0)} 
\tilde \Delta _{\alpha \uparrow }^{(\bar 0)*} + {\rm c.c.} ) 
\nonumber \\ && +B (
\tilde \Delta _{\bar \alpha \uparrow }^{(\bar 0)} 
\tilde \Delta _{\alpha \uparrow }^{(\bar 0)*}
+\tilde \Delta _{\bar \alpha \uparrow }^{(0)} 
\tilde \Delta _{\alpha \uparrow }^{(0)*} ) \Bigr \rbrack .
\label{F2}
\end{eqnarray}
$A_{\pm N}$ and $B$ are defined by (\ref{a12b}). 
We have introduced the new order parameters $\tilde \Delta _{\alpha \sigma
}^{(pN)}$ related to $\Delta _{\alpha \sigma }^{(pN)}$ by
\begin{eqnarray}
\tilde \Delta _{\alpha \sigma }^{(pN)}&=& I_{pN} (g_2 \Delta _{\alpha \sigma
}^{(pN)}+g_3 \Delta _{\bar \alpha \sigma }^{(\bar pN)})
= \tilde \Delta _{\bar \alpha \bar \sigma }^{(pN)*} , \nonumber \\ 
\Delta _{\alpha \sigma }^{(pN)} &=& 
\frac{g_2I_{\bar pN} \tilde \Delta _{\alpha \sigma }^{(pN)}
-g_3I_{pN}  \tilde \Delta _{\bar \alpha \sigma }^{(\bar pN)}}
{(g_2^2-g_3^2)I_{N} I_{\bar N}} ,
\end{eqnarray}
with $I_{pN}\equiv I_{pN}(Q_y^{(pN)})$. 

A commonly used approximation in the QNM is the quantum
limit approximation (QLA) valid when $\omega _c\gg
T$.\cite{rev,Virosztek86,Montambaux88} It consists in 
retaining only the most singular
(electron-hole) pairing channels that have the logarithmic singularity
$\sim \ln (2\gamma E_0/\pi T)$. This singularity results from 
pairings between electron and hole states of the same energy. Therefore, in the
QLA, $\tilde \Delta _{\alpha \sigma } (w_1,w_2)$ is nonzero only if
$\epsilon 
^{\bar \alpha }_{w_1\bar \sigma }=-\epsilon ^\alpha _{w_2\sigma }$. This
leads to 
\begin{eqnarray}
\tilde \Delta _{\alpha \sigma } (w_1,w_2) \Big \vert _{\rm QLA} &=& \sum
_{p=\pm } \delta 
_{k_{2x},k_{1x}+\alpha Q_x^{(pN)}} \delta _{l_2,l_1-\alpha pN} 
\nonumber \\ && \times 
e^{-i\alpha Q_y^{(pN)}b(2l_1-\alpha pN)/2} 
\tilde \Delta _{\alpha \sigma }^{(pN)} .
\label{paQLA}
\end{eqnarray}
The QLA is usually known as the single gap approximation (SGA) 
because it amounts to considering only the gap that opens at the Fermi
level. However, when umklapp processes are present, the spectrum cannot be 
described with only one gap, although gaps opening above and below the Fermi
level are still neglected (see section \ref{sec:spec}). For this reason, we
use the term QLA rather than SGA. In the QLA, we would have obtained Eq.\
(\ref{F2}), together with (\ref{a12b}), but with the exact susceptibility $\chi
^{(0)}_+$ replaced by  
\begin{equation} 
\chi ^{(0)}_+({\bf Q}_{pN}) \Big \vert _{\rm QLA}= I_{pN}^2 \frac{N(0)}{2} 
\ln \left ( \frac{2\gamma E_0}{\pi T} \right ) .
\end{equation}
Although qualitatively correct, this approximation strongly underestimates
the susceptibility $\chi ^{(0)}$ because it neglects terms $\sim \ln 
(2\gamma E_0/ \vert n\vert \omega _c)$ ($n\neq 0$) with respect to $\ln
(2\gamma E_0/\pi T)$. However, the QLA becomes very accurate for higher order
contributions to the free energy. For $F^{(4)}_N$, corrections to the QLA 
are of order $T^2/\omega _c^2$ and can therefore be neglected when $\omega
_c\gg T$. Within the QLA, we obtain
\bleq
\begin{eqnarray}
F^{(4)}_N &=& \frac{K}{2} \sum _\alpha \sum _{p_1,p_2,p_3,p_4}
\tilde \Delta _{\alpha \uparrow }^{(p_1N)}
\tilde \Delta _{\alpha \uparrow }^{(p_2N)*}
\tilde \Delta _{\alpha \uparrow }^{(p_3N)}
\tilde \Delta _{\alpha \uparrow }^{(p_4N)*} \nonumber \\ && \times
\exp \Bigl \lbrack iN(b/2) \lbrack p_1Q_y^{(p_1N)} - (2p_1-p_2)Q_y^{(p_2N)}
+(2p_1-2p_2+p_3)Q_y^{(p_3N)}-p_4Q_y^{(p_4N)}\rbrack \Bigr \rbrack
\nonumber \\ && \times
\delta _{(p_1-p_2+p_3-p_4)N,0} \sum _n \delta
_{Q_y^{(p_1N)}-Q_y^{(p_2N)}+Q_y^{(p_3N)} -Q_y^{(p_4N)},n2\pi/b} \,,
\label{fe4}
\end{eqnarray}
\eleq
where $K= 7\zeta (3)N(0)/(16\pi ^2T^2)$ and $\zeta (3)\simeq 1.20$.
A somewhat lengthly calculation (see appendix \ref{app:fe}) leads to 
\begin{eqnarray}
F^{(4)}_N &=& \frac{K}{2} \sum _{p,\alpha } 
\vert \tilde \Delta _{\alpha \uparrow }^{(pN)} \vert ^4 
+2K \sum _\alpha \vert \tilde \Delta _{\alpha \uparrow }^{(N)}
\tilde \Delta _{\alpha \uparrow }^{(\bar N)} \vert ^2 +\delta F_0^{(4)}
\nonumber \\
\delta F_0^{(4)} &=& 
\delta _{N,0} \Bigl ( \sum _n \delta _{Q_y^{(0)}b,n\pi } \Bigr ) K
\sum _\alpha \Bigl \lbrack \tilde \Delta _{\alpha \uparrow }^{(0)}
\vert \tilde \Delta _{\alpha \uparrow }^{(\bar 0)} \vert ^2 
\tilde \Delta _{\alpha \uparrow }^{(\bar 0)*} \nonumber \\ && 
+ \tilde \Delta _{\alpha \uparrow }^{(\bar 0)}
\vert \tilde \Delta _{\alpha \uparrow }^{(0)} \vert ^2 
\tilde \Delta _{\alpha \uparrow }^{(0)*}
+\frac{1}{2}(\tilde \Delta _{\alpha \uparrow }^{(0)}
\tilde \Delta _{\alpha \uparrow }^{(\bar 0)*})^2 + {\rm c.c.} \Bigr \rbrack .
\nonumber \\  
\label{F4}
\end{eqnarray}
Eqs.\ (\ref{F2}) and (\ref{F4}) show that the case $N=0$ and $Q_y^{(0)}=0,\pi
/b$ is special, since in that case one cannot distinguish between $\Delta
_{\alpha \uparrow }^{(0)}$ and $\Delta _{\alpha \uparrow }^{(\bar 0)}$. In
practice, the case $Q_y^{(0)}=0$ never occurs (see section \ref{sec:imp}) so
that we shall not consider it any more.

\subsection{Phases $N\neq 0$, or $N=0$ and $Q_y^{(0)}\neq \pi /b$ }
\label{subsec:polarA}

We first consider the cases $N\neq 0$, and  $N=0$ with $Q_y^{(0)}\neq \pi
/b$. The latter corresponds to the reentrant phases $N=0$ within the cascade 
(see section \ref{sec:imp}). The quadratic part of the free energy
(\ref{F2}) is not diagonal in the order parameters since $\tilde \Delta
_{\alpha 
\uparrow }^{(N)}$ is coupled to $\tilde \Delta _{\bar \alpha \uparrow }^{(\bar
N)}$.  Introducing the new order parameters $u^{(N)}_\alpha $ and
$v^{(N)}_\alpha $ defined by 
\begin{equation}
\left (
\begin{array}{l}
\tilde \Delta _{\alpha \uparrow }^{(N)} \\ 
\tilde \Delta _{\bar \alpha \uparrow }^{(\bar N)} 
\end{array} \right )
= \left (
\begin{array}{lr}
\cos (\vartheta _N) & 
-\sin (\vartheta _N) \\
\sin (\vartheta _N) & 
\cos (\vartheta _N) \\
\end{array}
\right )   
\left (
\begin{array}{l}
u_\alpha ^{(N)} \\ v_\alpha ^{(N)}
\end{array} \right ),
\label{chgt}
\end{equation}
we obtain 
\begin{eqnarray}
F^{(2)}_N&=& \sum _\alpha (\tilde \Delta _{\alpha \uparrow }^{(N)*}, \,\,\,\,\,
\tilde \Delta _{\bar \alpha \uparrow }^{(\bar N)*})
\left (
\begin{array}{lr}
A_N  & B \\ B  & A_{\bar N} \\
\end{array}
\right )
\left (
\begin{array}{l}
\tilde \Delta _{\alpha \uparrow }^{(N)} \\ 
\tilde \Delta _{\bar \alpha \uparrow }^{(\bar N)} 
\end{array} \right )
\nonumber \\ &=& 
\sum _\alpha \left \lbrack \lambda _1^{(N)} \vert u_\alpha ^{(N)}\vert ^2 
+ \lambda _2^{(N)} \vert v_\alpha ^{(N)}\vert ^2 \right \rbrack 
\end{eqnarray}
for 
\begin{equation}
\tan (2\vartheta _N)= \frac{2B }{A_N-A_{\bar N}} .
\end{equation}
$2\vartheta _N$ can be chosen in $\rbrack -\pi /2,\pi /2\rbrack $.
The eigenvalues $\lambda _1^{(N)}$ and $\lambda _2^{(N)}$ are given by
\begin{eqnarray}
\lambda _1^{(N)} &=& A_N \cos ^2 (\vartheta _N)+A_{\bar N} \sin ^2
(\vartheta _N)+B\sin  (2\vartheta _N) 
\nonumber \\ 
&=& \frac{A_N+A_{\bar N}}{2} 
+\frac{1}{2}{\rm sgn}(A_N-A_{\bar N})\nonumber \\ && \times 
\lbrack (A_N-A_{\bar N})^2+4B^2 \rbrack ^{1/2} , \nonumber \\
\lambda _2^{(N)} &=& A_N \sin ^2(\vartheta _N)+A_{\bar N} \cos ^2(\vartheta _N)-B\sin
(2\vartheta _N)  
\nonumber \\ 
&=& \frac{A_N+A_{\bar N}}{2}
-\frac{1}{2}{\rm sgn}(A_N-A_{\bar N})\nonumber \\ && \times 
\lbrack (A_N-A_{\bar N})^2+4B^2 \rbrack ^{1/2} .
\end{eqnarray}
The transition temperature $T_c^{(N)}$ is determined by ${\rm min}(\lambda
_1^{(N)},\lambda _2^{(N)})=0$. In this section, the sign of $N$ has not been
specified yet,
since both phases $N$ and $-N$ correspond to a phase with two SDWs at
wave vectors $(Q_x^{(N)},Q_y^{(N)})$ and $(Q_x^{(\bar N)},-Q_y^{(N)})$. In
the following, 
we fix the sign of $N$ by the condition $\chi ^{(0)}_+({\bf Q}_N)\geq \chi
^{(0)}_+({\bf Q}_{\bar N})$ (i.e., in the absence of umklapp processes, the
phase 
$N$ would be more stable than the phase $-N$). Then we have $0\leq A_N \leq
A_{\bar N}$ and $0\leq \lambda _1^{(N)}\leq \lambda _2^{(N)}$ for $T\geq
T_c^{(N)}$ 
(assuming $0\le g_3<g_2$). The transition temperature $T_c^{(N)}$ is 
determined by $\lambda _1^{(N)}=0$,    
i.e., $A_NA_{\bar N}=B^2$. Using Eqs.\ (\ref{a12b}), we recover the result
(\ref{stoner}) obtained from the RPA calculation of the susceptibility. 
 
Since $\lambda _2^{(N)}>0$ for $T\lesssim T_c^{(N)}$, we can put $v_\alpha
^{(N)}=0$ in 
the expansion of the free energy. This leads to 
\begin{eqnarray}
F_N &=& \lambda _1^{(N)} \sum _\alpha \vert u_\alpha ^{(N)}\vert ^2 
+\frac{K}{2}\lbrack \cos ^4(\vartheta _N)+\sin ^4(\vartheta _N)\rbrack  
\nonumber \\ && \times 
\sum _\alpha \vert u_\alpha ^{(N)}\vert ^4
+K \sin ^2(2\vartheta _N)
\vert u_+^{(N)} u_-^{(N)} \vert ^2 .
\label{fe1}
\end{eqnarray}
The minimum of the free energy is determined by 
\begin{eqnarray}
\frac{\partial F_N}{\partial u_\alpha ^{(N)*}} &=&
u_\alpha ^{(N)} \Bigl \lbrack
\lambda _1^{(N)} +K
\lbrack \cos ^4(\vartheta _N)+\sin ^4(\vartheta _N)\rbrack 
\vert u_\alpha ^{(N)}\vert ^2 \nonumber \\ && 
+K \sin ^2(2\vartheta _N)
\vert u_{\bar \alpha }^{(N)}\vert ^2 \Bigr \rbrack =0. 
\end{eqnarray}
For $\lambda _1^{(N)}<0$ (i.e., $T<T_c^{(N)}$), there are  two solutions
corresponding to a SDW phase: 

i) $\vert u_+^{(N)}\vert =\vert u_-^{(N)}\vert $
($\neq 0$). Using $v_\alpha ^{(N)}=0$, we find $\vert \tilde \Delta
_{+\uparrow }^{(N)}\vert = \vert \tilde \Delta _{-\uparrow }
^{(N)}\vert $, $\vert \tilde \Delta _{+\uparrow }^{(\bar N)}\vert = \vert
\tilde \Delta _{-\uparrow }^{(\bar N)}\vert $, and also
$\vert \Delta _{+\uparrow }^{(N)}\vert = \vert \Delta _{-\uparrow
}^{(N)}\vert $, 
$\vert \Delta _{+\uparrow }^{(\bar N)}\vert = \vert \Delta _{-\uparrow }^{(\bar
N)}\vert $. From the results of section \ref{sec:op}, we conclude that 
the two SDWs are linearly polarized. 

ii) $u_+^{(N)}\neq 0$ and $u_-^{(N)}=0$
(or the symmetric solution: $u_-^{(N)}\neq 0$ and $u_+^{(N)}=0$). $v_\alpha
^{(N)}=0$ then implies $\tilde \Delta _{-\uparrow }^{(N)}=
\tilde \Delta _{+\uparrow }^{(\bar N)}=0$, and $\Delta _{-\uparrow }^{(N)}=
\Delta _{+\uparrow }^{(\bar N)}=0$. This corresponds to two helicoidal SDWs
of opposite chiralities. 
 
\begin{figure}
\epsfxsize 15cm 
\epsffile[50 80 600 740]{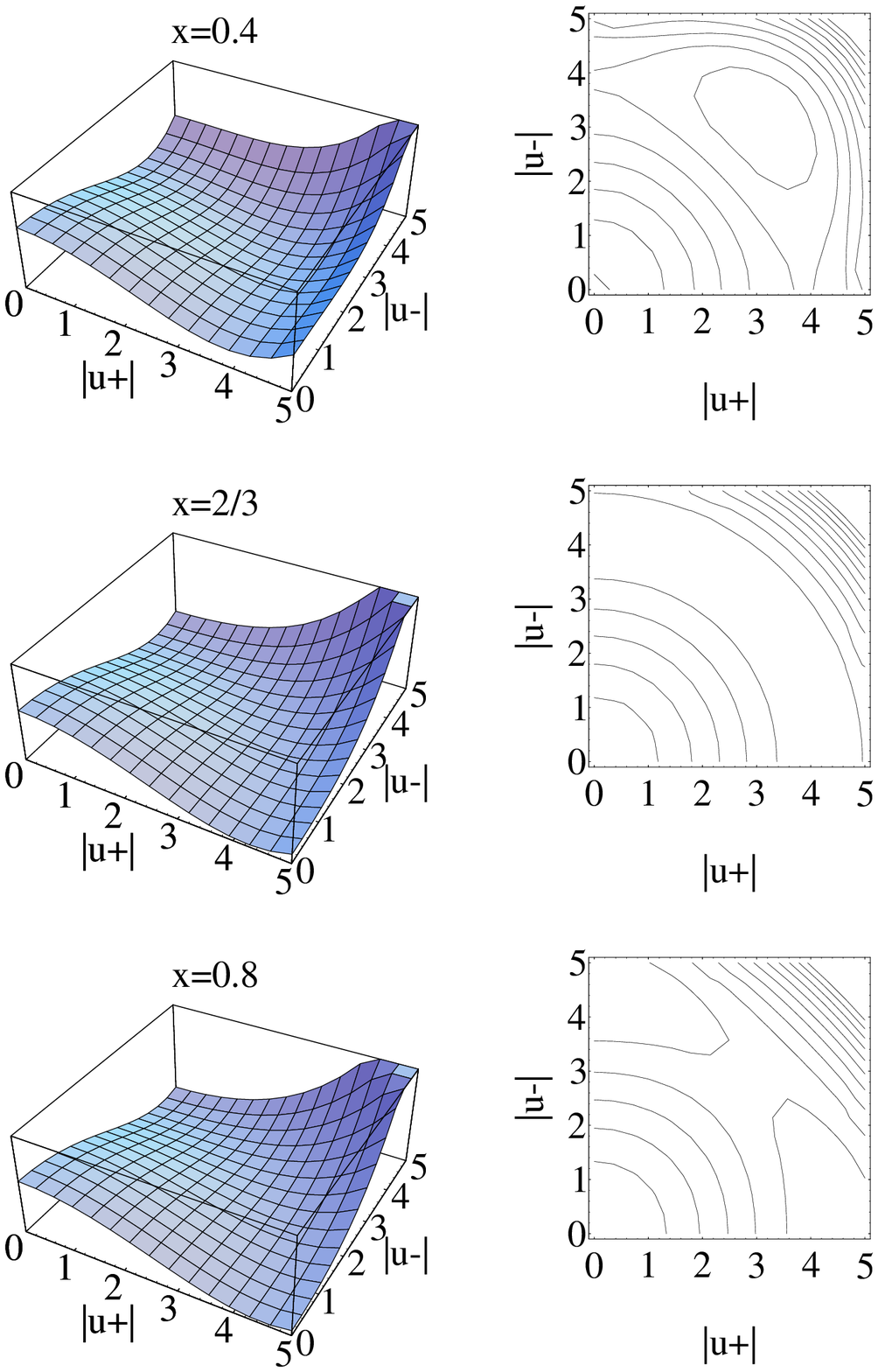}
\refstepcounter{figure} \label{Fig:energy}
{\small FIG. \ref{Fig:energy}.
Free energy $F_N$ as a function of $\vert u^{(N)}_+\vert $ and $\vert
u^{(N)}_-\vert $ for different values of the parameter $x=\sin ^2(2\vartheta
_N)$. When $x<2/3$, the minimum of $F_N$ corresponds to $\vert
u^{(N)}_+\vert =\vert u^{(N)}_- \vert $ (sinusoidal phase). When $x>2/3$,
there are two minima located on the line $ u^{(N)}_+=0$ and $u_-^{(N)}=0$
(helicoidal phase). When $x=2/3$, the minima are infinitively degenerate. }
\end{figure}

For the sinusoidal waves (solution (i)), we find 
\begin{eqnarray}
\vert u^{(N)}_\alpha \vert ^2 &=& -
\frac{\lambda _1^{(N)}}{K \left (1+\frac{1}{2}\sin ^2
(2\vartheta _N)\right )} , \label{poS} \\ 
F_N &=& - \frac{{\lambda _1^{(N)}}^2}{K \left (1+\frac{1}{2}\sin ^2
(2\vartheta _N)\right )} ,
\label{feS}
\end{eqnarray}
while for the helicoidal waves (solution (ii)), we have 
\begin{eqnarray}
\vert u^{(N)}_+\vert ^2 &=& -
\frac{\lambda _1^{(N)}}{K
\left (1-\frac{1}{2}\sin ^2 (2\vartheta _N)\right ) } ,
\,\,\,\,\, u_-^{(N)} = 0 , 
\label{poH} \\ 
F_N &=& - \frac{{\lambda _1^{(N)}}^2}{2K \left (1-\frac{1}{2}\sin ^2
(2\vartheta _N)\right )} .
\label{feH}
\end{eqnarray}
Comparing Eqs.\ (\ref{feS}) and (\ref{feH}), we conclude that the
helicoidal phase is more stable than the sinusoidal phase when $\sin
^2(2\vartheta _N)> 2/3$, i.e., when $\sqrt{2}|B|>|A_N-A_{\bar N}|$. The free
energy $F_N$ (Eq.\ (\ref{fe1})) is shown in Fig.\ \ref{Fig:energy} as a 
function of $\vert u_+^{(N)}\vert $ and $\vert u_-^{(N)}\vert $ for $\lambda
_1^{(N)}<0$. When $\sin ^2(2\vartheta _N)< 2/3$, the minimum of $F_N$
corresponds to  
$\vert u_+^{(N)}\vert =\vert u_-^{(N)}\vert $. When $\sin ^2(2\vartheta
_N)>2/3$, there 
are two minima located on the lines $u_+^{(N)}=0$ and $u_-^{(N)}=0$. For $\sin
^2(2\vartheta _N)=2/3$, the minima are infinitively degenerate. 

In the absence of umklapp processes ($g_3=0$), $\vartheta _N=0$ and the SDWs
are sinusoidal. For values of $g_3$ such that $Q_y^{(N)}=\pi /b$, the
polarization of the waves depends on the details of the Fermi surface. In
our model, it is determined by $t_{4b}$  (see the numerical
calculation of section \ref{sec:imp}). 

The results obtained in this section are very simple when $N=0$. 
Since $\chi ^{(0)}_+({\bf Q}_{N})=\chi ^{(0)}_+({\bf Q}_{\bar N})$ for
$N=0$ (the susceptibility does not 
change when $Q_y\to -Q_y$), $A_N=A_{\bar N}$ and $\vartheta _0=\pi /4$. The
phase $N=0$ (with $Q_y\ne \pi /b$) is therefore always helicoidal. This result
does not hold when $Q_y=\pi /b$ (see section \ref{subsec:polarN0}).

Using $v_\alpha ^{(N)}=0$, we obtain the ratio of the amplitudes of the two
SDWs: 
\begin{eqnarray}
\tilde \gamma &=& \frac{\tilde \Delta _{\bar \alpha \uparrow }^{(\bar
N)}}{\tilde \Delta _{\alpha \uparrow }^{(N)}} = \tan (\vartheta _N) ,
\label{gamt} \\ 
\gamma &=& \frac{\Delta _{\bar \alpha \uparrow }^{(\bar N)}}{\Delta
_{\alpha \uparrow }^{(N)}} = \frac{\tilde \gamma I_N-rI_{\bar N}}{I_{\bar
N}-r\tilde \gamma I_N} . 
\end{eqnarray}
Since $\vartheta _N\in \rbrack -\pi /4,\pi /4\rbrack $, $\vert \gamma \vert ,
\vert \tilde \gamma \vert \le 1$. The SDW with wave vector ${\bf Q}_N$
has the largest amplitude. This simply results from the fact that we have
chosen $N$ such that $\chi ^{(0)}({\bf Q}_N)\ge \chi ^{(0)}({\bf Q}_{\bar
N})$. The transition from the sinusoidal phase to the helicoidal phase
occurs when $\sin ^2(2\vartheta _N)=2/3$, i.e., when $\vert \tilde \gamma \vert
= 0.518$. 

Note that in the positive phases, it is $\tilde \gamma $ and not
$\gamma $ which is vanishingly small. When $\chi ^{(0)}_+({\bf Q}_{\bar N})
\ll \chi ^{(0)}_+({\bf Q}_{N})$, the minimum of the free energy corresponds
to $\tilde \Delta _{\alpha \uparrow }^{(\bar N)}=0$ (and not $\Delta
_{\alpha \uparrow }^{(\bar N)}=0$), since it is the potential $\tilde \Delta
_{\alpha \uparrow }$ (and not $\Delta _{\alpha \uparrow }$) that couples to
the electrons (see Eq.\ (\ref{ham2})). $\tilde \gamma \simeq 0$ implies
$|\gamma | \simeq r$.

\subsection{Phase $N=0$, $Q_y^{(0)}=\pi /b$} 
\label{subsec:polarN0}

We now consider the case $N=0$ with $Q_y^{(0)}=\pi /b$, which corresponds to
the phase $N=0$ terminating the cascade. There is only one SDW in
that case, so we can choose $\Delta _{\alpha \sigma }^{(0)}= \Delta
_{\alpha \sigma }^{(\bar 0)}$ (alternatively, we could choose $\Delta
_{\alpha \sigma }^{(\bar 0)}=0$). The free energy is then given by
\begin{equation}
F_0 = 4 \sum _\alpha \Bigl \lbrack A_0
\vert \tilde \Delta _{\alpha \uparrow }^{(0)} \vert ^2 +B
 \tilde \Delta _{\alpha \uparrow }^{(0)}
\tilde \Delta _{\bar \alpha \uparrow }^{(0)*} 
+2K \vert \tilde \Delta _{\alpha \uparrow }^{(0)} \vert
^4 \Bigr \rbrack  .
\label{feN0}
\end{equation}
The quadratic part of the free energy $F_0^{(2)}$ is diagonalized by
introducing the order parameters $u$ and $v$ defined by
\begin{equation}
\left (
\begin{array}{l}
\tilde \Delta _{+\uparrow }^{(0)} \\ 
\tilde \Delta _{-\uparrow }^{(0)} 
\end{array} \right )
= \frac{1}{\sqrt{2}} 
\left (
\begin{array}{lr}
1 & -1 \\ 1 & 1 \\
\end{array}
\right )
\left (
\begin{array}{l}
u \\ v
\end{array} \right ) .
\end{equation}
This leads to   
\begin{equation}
F_0^{(2)} = 4 \lambda _1^{(0)}\vert u\vert ^2 +4 \lambda _2^{(0)}\vert
v\vert ^2 , 
\end{equation}
with
\begin{eqnarray}
\lambda _1^{(0)} &=& \frac{1}{I_0^2} \left ( \frac{1}{g_2+g_3}-\chi
^{(0)}_+({\bf Q}_0) \right ) , \nonumber \\ 
\lambda _2^{(0)} &=& \frac{1}{I_0^2} \left ( \frac{1}{g_2-g_3}-\chi
^{(0)}_+({\bf Q}_0) \right ) .
\end{eqnarray}
$T_c^{(0)}$ is determined by $\lambda _1^{(0)}=0$, i.e., $1-(g_2+g_3)\chi
^{(0)}_+({\bf Q}_0)=0$. This agrees with the RPA result (\ref{stoner}) when 
$\chi ^{(0)}_+({\bf Q}_N)=\chi ^{(0)}_+({\bf Q}_{\bar N})=\chi ^{(0)}_+({\bf
Q}_0)$.  

Since $\lambda _2^{(0)}>\lambda _1^{(0)}$, $v=0$ for $T\lesssim
T_c^{(0)}$. This implies 
$\Delta _{+\uparrow }^{(0)}=\Delta _{-\uparrow }^{(0)}$. Thus, the phase $N=0$,
$Q_y=\pi /b$  is
always linearly polarized. For $N=0$ (and $Q_y^{(0)}=\pi /b$), $g_3$ couples 
$\tilde \Delta _{+\uparrow }^{(0)}$ and $\tilde \Delta _{-\uparrow }
^{(0)}$. As a result, we cannot have $\tilde \Delta _{+\uparrow }^{(0)}\neq
0$ and $\tilde 
\Delta _{-\uparrow }^{(0)}=0$, so that the polarization cannot be circular. The
case $g_3=0$ is special since then $\lambda _1^{(0)}=\lambda _2^{(0)}$ and
both $u$ 
and $v$ become nonzero below $T_c^{(0)}$. Nevertheless, Eq.\ (\ref{feN0})
shows that the sinusoidal wave is more stable than the helicoidal wave in this
case too. 

Setting $v=0$, we obtain the free energy 
\begin{equation}
F_0=4\lambda _1^{(0)} \vert u\vert ^2 +4K \vert u\vert ^4 .
\end{equation}
Minimizing $F_0$ with respect to $u$, we obtain 
\begin{equation}
\vert u\vert ^2= - \frac{\lambda _1^{(0)}}{2K} , \,\,\,\,\,
F_0 = -\frac{(\lambda _1^{(0)})^2}{K} .
\end{equation}

\section{spectrum}
\label{sec:spec}

In the absence of umklapp processes, the spectrum consists of Landau
subbands separated by gaps opening at the Fermi level, and at $n\omega
_c/2$ ($n$ integer) above and below the Fermi
level.\cite{rev,Yamaji85,Virosztek86,Montambaux88} We show in this section
that for sinusoidal SDWs the gaps opening at the Fermi level depend on the
transverse momentum $k_y$ (in this section, we use a gauge where $k_y$ is a
good quantum number). On the other hand, the spectrum of the
helicoidal phase remains independent of $k_y$. As shown
below, this provides a natural explanation of the instability of the
sinusoidal phase with respect to the helicoidal phase. 

In this section we use the gauge ${\bf A}=(0,Hx,0)$ where the calculations
are simpler. In the next section this will also allow us to calculate the
QHE using previous results derived in this gauge. Using the Peierls
substitution, we obtain 
\begin{eqnarray}
{\cal H}_0 &=& \sum _{\alpha ,\sigma ,k_y} \int dx \,
\psi _{\alpha \sigma }^\dagger (x,k_y) 
\lbrack v_F(-i\alpha \partial _x -k_F) \nonumber \\ && 
+t_\perp (k_yb-Gx) +\sigma \mu _BH 
\rbrack \psi _{\alpha \sigma } (x,k_y) ,
\end{eqnarray}
where we have taken the Fourier transform with respect to the $y$ direction. 
In the absence of electron-electron interaction, the eigenstates and the
spectrum are given by 
\begin{eqnarray}
\phi _{\bf k}^\alpha ({\bf r}) &=& \frac{1}{\sqrt{S}} e^{i{\bf k}\cdot {\bf
r}+i\frac{\alpha }{\omega _c} T_\perp (k_yb-Gx)} , \nonumber \\
\epsilon _{{\bf k},\sigma }^\alpha &=& v_F(\alpha k_x - k_F)+\sigma \mu _BH .
\end{eqnarray}
The dispersion
law is now one-dimensional and the states are extended in both the $x$ and
$y$ directions. However, since the energy does not depend on $k_y$, we can
take the Fourier transform with respect to $k_y$ and obtain localized wave
functions similar to those obtained in section \ref{sec:imp}. The
Wannier-Stark ladder can then be recovered by a simple gauge
transformation. 

Using (\ref{ham2}), the interacting part of the mean-field Hamiltonian can be
written as
\bleq
\begin{eqnarray}
{\cal H}_{\rm int}^{\rm MF} &=& -\sum _{\alpha ,k_y} \int dx 
\Biggl \lbrack e^{-i\alpha Q_x^{(N)}x} \frac{\tilde \Delta _{\alpha \uparrow
}^{(N) 
}}{I_N} \psi _{\bar \alpha \downarrow }^\dagger (x,k_y)
\psi _{\alpha \uparrow } (x,k_y+\alpha Q_y^{(N)}) 
\nonumber \\ &&  +
e^{i\alpha Q_x^{(\bar N)}x} \frac{\tilde \Delta _{\bar \alpha \uparrow
}^{(\bar N)}}{I_{\bar N}} \psi _{\alpha \downarrow }^\dagger (x,k_y)
\psi _{\bar \alpha \uparrow }(x,k_y+\alpha Q_y^{(N)}) 
\Biggr \rbrack +{\rm h.c.}
\end{eqnarray}
up to a constant term. Introducing new fields 
$\tilde \psi _{\alpha \sigma }^{(\dagger )}$ defined by
\begin{equation}
\psi _{\alpha \sigma } (x,k_y)=
 e^{ i\alpha k_Fx-i\alpha \sigma \frac{\mu
_BH}{v_F} x +i\frac{\alpha }{\omega _c} T_\perp (k_yb-Gx) } 
\tilde \psi _{\alpha \sigma } (x,k_y) , 
\end{equation}
we rewrite the Hamiltonian as
\begin{eqnarray}
{\cal H}_{\rm MF} &=& \sum _{\alpha ,\sigma ,k_y} \int dx \,
\tilde \psi _{\alpha \sigma }^\dagger (x,k_y) 
(-i\alpha v_F\partial _x) 
\tilde \psi _{\alpha \sigma } (x,k_y) -\sum _{\alpha ,k_y,n} 
\int dx\, e^{-i\alpha NGx -i\alpha n(k_yb-Gx+\alpha Q_y^{(N)}b/2)} 
\nonumber \\ && \times 
\Biggl \lbrack \frac{\tilde \Delta _{\alpha \uparrow }^{(N)}}{I_N}
I_n (Q_y^{(N)})
\tilde \psi _{\bar \alpha \downarrow }^\dagger (x,k_y)
\tilde \psi _{\alpha \uparrow }(x,k_y+\alpha Q_y^{(N)}) 
+ \frac{\tilde \Delta _{\bar \alpha \uparrow }^{(\bar N)}}{I_{\bar N}}
I_{-n} (Q_y^{(N)})
 \tilde \psi _{\alpha \downarrow }^\dagger (x,k_y)
\tilde \psi _{\bar \alpha \uparrow }(x,k_y+\alpha Q_y^{(N)})  +{\rm h.c.}
\Biggr \rbrack .  \nonumber \\
\label{ham3} 
\end{eqnarray}
We have used
\begin{equation}
\sum _{n=-\infty }^\infty I_n(q_y) e^{-in(u+q_yb/2)} = 
e^{\frac{i}{\omega _c} \lbrack T_\perp (u)+T_\perp (u+q_yb) \rbrack } ,
\end{equation}
which follows from (\ref{Idef}). 
To proceed further, we use the QLA. This amounts to retaining only the gaps
that open at the Fermi level neglecting those opening above and below the
Fermi level. In the QLA, only the term $n=N$ is retained in
(\ref{ham3}). This leads to 
\begin{eqnarray}
{\cal H}_{\rm int}^{\rm MF} &=& -\sum _{\alpha ,k_y} \int dx \,
e^{-i\alpha Nk_yb} \Bigl \lbrack 
\tilde \Delta _{\alpha \uparrow }^{(N)} 
\tilde \psi _{\bar \alpha \downarrow }^\dagger (x,k_y-\alpha Q_y^{(N)}/2)
\tilde \psi _{\alpha \uparrow }(x,k_y+\alpha Q_y^{(N)}/2)
\nonumber \\ && 
+\tilde \Delta _{\bar \alpha \uparrow }^{(\bar N)} 
\tilde \psi _{\alpha \downarrow }^\dagger (x,k_y-\alpha Q_y^{(N)}/2)
\tilde \psi _{\bar \alpha \uparrow }(x,k_y+\alpha Q_y^{(N)}/2)
\Biggr \rbrack +{\rm h.c.}
\label{ham4}
\end{eqnarray}

In the case of sinusoidal SDWs, Eq.\ (\ref{ham4}) shows that $k_y$ is
coupled to $k_y\pm Q_y^{(N)}$, $k_y\pm 2Q_y^{(N)}$,... so that the Hamiltonian
cannot be straightforwardly diagonalized. For very small $g_3$, the system
is not significantly affected by umklapp processes (the SDW with wave vector
${\bf Q}_{\bar N}$ has a very small amplitude), 
and therefore we expect that the spectrum will not be very
sensitive to $g_3$. Thus, the most interesting situation arises when 
umklapp processes are strong enough so that the FISDW phase becomes
commensurate in the transverse direction ($Q_y^{(N)}=\pi /b$) (which implies
that $N$ is even). In that case, $k_y$ is coupled only to $k_y+\pi /b$, and
both the sinusoidal and helicoidal phases are described by the Hamiltonian
\begin{eqnarray}
{\cal H}_{\rm MF} &=& \sum _{K_x,k_y} 
\Bigl (\tilde \psi _{+\uparrow }^\dagger (K_x,k_y+\pi /2b), \,\,
\tilde \psi _{-\downarrow }^\dagger (K_x,k_y-\pi /2b) \Bigr ) 
\nonumber \\ && \times 
\left ( 
\begin{array}{lr}
v_FK_x  & -e^{iNk_yb} \tilde \Delta _{+\uparrow }^{(N)*} 
-e^{-iNk_yb} \tilde \Delta _{+\uparrow }^{(\bar N)*} \\ 
-e^{-iNk_yb} \tilde \Delta _{+\uparrow }^{(N)} 
-e^{iNk_yb} \tilde \Delta _{+\uparrow }^{(\bar N)} 
& -v_FK_x \\
\end{array}
\right )
\left (
\begin{array}{l}
\tilde \psi _{+\uparrow }(K_x,k_y+\pi /2b) \\ 
\tilde \psi _{-\downarrow }(K_x,k_y-\pi /2b)
\end{array} \right )
\nonumber \\  && + \sum _{K_x,k_y}
\Bigl (\tilde \psi _{+\downarrow }^\dagger (K_x,k_y-\pi /2b), \,\,
\tilde \psi _{-\uparrow }^\dagger (K_x,k_y+\pi /2b)\Bigr ) 
\nonumber \\ && \times 
\left ( 
\begin{array}{lr}
v_FK_x  & -e^{-iNk_yb} \tilde \Delta _{-\uparrow }^{(\bar N)} 
-e^{iNk_yb} \tilde \Delta _{-\uparrow }^{(N)} \\ 
-e^{iNk_yb} \tilde \Delta _{-\uparrow }^{(\bar N)*} 
-e^{-iNk_yb} \tilde \Delta _{-\uparrow }^{(N)*}  
& -v_FK_x \\
\end{array}
\right )
\left (
\begin{array}{l}
\tilde \psi _{+\downarrow } (K_x,k_y-\pi /2b) \\
\tilde \psi _{-\uparrow }(K_x,k_y+\pi /2b)
\end{array} \right ) ,
\label{ham5}
\end{eqnarray}
\eleq
where $K_x$ is now measured with respect to $\pm k_F\pm \mu _BH/v_F$.
In the metallic phase, the dispersion law is given by $\epsilon _1=\pm
v_FK_x$, $\epsilon _2=\pm v_FK_x$. The $+/-$ corresponds to right/left
moving electrons and the degeneracy is due to spin. In the SDW
phase, gaps open at the Fermi level, and the dispersion law becomes
$E_1=\pm \lbrack v_F^2K_x^2 + \Delta _1(k_y)^2\rbrack ^{1/2}$ and $E_2=\pm
\lbrack v_F^2K_x^2 + \Delta _2(k_y)^2\rbrack ^{1/2}$ with
\begin{eqnarray}
\Delta _1(k_y)^2 &=& \vert e^{-iNk_yb} \tilde \Delta _{+\uparrow }^{(N)} +
 e^{iNk_yb} \tilde \Delta _{+\uparrow }^{(\bar N)} \vert ^2  \nonumber \\
&=& \vert \tilde \Delta _{+\uparrow }^{(N)} \vert ^2 + 
\vert \tilde \Delta _{+\uparrow }^{(\bar N)} \vert ^2 \nonumber \\ &&
+ 2\vert \tilde \Delta _{+\uparrow }^{(N)} \tilde \Delta _{+\uparrow
 }^{(\bar N)} 
\vert \cos (2Nk_yb+\varphi _1) , \nonumber \\
\Delta _2(k_y)^2 &=& \vert e^{iNk_yb} \tilde \Delta _{-\uparrow }^{(N)} +
 e^{-iNk_yb} \tilde \Delta _{-\uparrow }^{(\bar N)} \vert ^2 \nonumber \\
&=& \vert \tilde \Delta _{-\uparrow }^{(N)} \vert ^2 +
\vert \tilde \Delta _{-\uparrow }^{(\bar N)} \vert ^2 \nonumber \\ &&
+2\vert \tilde \Delta _{-\uparrow }^{(N)} \tilde \Delta _{-\uparrow }^{(\bar
 N)} 
\vert \cos (2Nk_yb+\varphi _2) ,
\label{gap12}
\end{eqnarray}
where $\varphi _1$ and $\varphi _2$ depend on the phases of $\tilde \Delta
_{\alpha \uparrow }^{(pN)}$. 

\subsection{Sinusoidal waves}

For sinusoidal waves, as follows from (\ref{gap12}), there are interferences
between the two SDWs, and
the dispersion depends on $k_y$. Using $\vert \tilde \Delta _{+\uparrow
}^{(\bar N)}\vert /\vert \tilde \Delta _{+\uparrow }^{(N)}\vert = \vert \tan
(\vartheta _N) \vert $ (Eq.\ (\ref{gamt})), we obtain
\begin{equation}
\frac{\vert \Delta _1(k_y)\vert _{\rm min}}
{\vert \Delta _1(k_y)\vert _{\rm max}} = 
\frac{\vert \Delta _2(k_y)\vert _{\rm min}}
{\vert \Delta _2(k_y)\vert _{\rm max}} = 
\frac{1-\vert \tan (\vartheta _N) \vert }{1+\vert \tan (\vartheta _N) \vert } .
\label{rap}
\end{equation}
For $\vert \vartheta _N\vert \ll 1$, which corresponds to a large $t_{4b}$
(section \ref{sec:imp}), $\vert \tilde \Delta _{+\uparrow }^{(\bar N)}\vert
\ll \vert \tilde \Delta _{+\uparrow }^{(N)} \vert $. The dispersion in
the transverse direction is weak. Everywhere on the Fermi surface, the gap
is of the order of $\vert \tilde \Delta _{+\uparrow }^{(N)}\vert =
\vert \tilde \Delta _{-\uparrow }^{(N)} \vert $. 
For larger values of $\vert \vartheta _N\vert $, the dispersion becomes
significant. In particular, when $\vert \vartheta _N\vert =\pi /4$ (which
occurs when there is a degeneracy between $N$ and $-N$, i.e., when
$t_{4b}=0$), we have $\vert
\tilde \Delta _{+\uparrow }^{(N)}\vert =  \vert \tilde \Delta _{+\uparrow
}^{(\bar N)}\vert $ and the spectrum becomes gapless. \cite{Lebed91a} Eqs.\
(\ref{gap12}) show that there are $4N$ values of $k_y$ for which $\Delta _1$
or $\Delta _2$ vanish.  

However, the occurrence of the helicoidal phase prevents the spectrum from
becoming gapless. Indeed, the stability of the sinusoidal phase requires $\sin
^2(2\vartheta _N)<2/3$. From (\ref{rap}), we then deduce 
\begin{equation}
\frac{\vert \Delta _1(k_y)\vert _{\rm min}}
{\vert \Delta _1(k_y)\vert _{\rm max}} \ge  0.32 .
\end{equation}

\subsection{Helicoidal waves}

For helicoidal waves, $\tilde \Delta _{-\uparrow }^{(N)}= \tilde \Delta
_{+\uparrow }^{(\bar N)}=0$. The spectrum is given by $E_1=\pm \lbrack
v_F^2K_x^2+\vert \tilde \Delta _{+\uparrow }^{(N)}\vert ^2 \rbrack ^{1/2}$ and
$E_2=\pm \lbrack v_F^2K_x^2+\vert \tilde \Delta _{-\uparrow }^{(\bar N)}\vert
^2\rbrack ^{1/2}$. Contrary to the sinusoidal phase, the gaps at the Fermi
level, $\tilde \Delta _{+\uparrow }^{(N)}$ and $\tilde \Delta _{-\uparrow
}^{(\bar N)}$, remain independent of the transverse momentum $k_y$. 
Since the sinusoidal phase becomes
gapless at $\vert \vartheta _N\vert =\pi /4$, it is natural to expect that,
above a certain value of $\vert \vartheta _N\vert $, the system prefers to form
helicoidal SDWs in 
order to lower the free energy by opening a large gap on the whole Fermi
surface. This is precisely the result we have obtained in section
\ref{sec:polar} by considering the Ginzburg-Landau expansion of the free
energy. 

In general, the two gaps $\vert \tilde \Delta _{+\uparrow }^{(N)}\vert $ and
$\vert \tilde \Delta _{-\uparrow }^{(\bar N)}\vert $ differ (they are equal
only when $t_{4b}=0$ or $N=0$). As shown in section \ref{sec:mee}, this
property gives rise to a kinetic magnetoelectric effect.

\section{Quantum Hall Effect}
\label{sec:qhe}

In this section, we use the results of Ref.\ \onlinecite{Yakovenko91} to
study the quantum Hall effect. At zero temperature, the offdiagonal
conductivity is given by the formula 
\begin{eqnarray}
\sigma _{xy} &=& -ie^2 \sum _a \int \frac{dk_x}{2\pi } \frac{dk_y}{2\pi } 
\bigl ( \partial _{k_x} \langle \psi _a \vert \partial _{k_y} \vert 
\psi _a \rangle \nonumber \\ &&
- \partial _{k_y} \langle \psi _a \vert \partial _{k_x} 
\vert \psi _a \rangle \Bigr ) .
\label{Sxy}
\end{eqnarray}
The summation is taken over all completely occupied bands and the integral
is taken over the Brillouin zone. $\vert \psi _a (k_x,k_y)
\rangle $ are the normalized eigenvectors of the Hamiltonian. 

Before calculating $\sigma _{xy}$ in the FISDW phases, we recall the main
result of Ref.\ \onlinecite{Yakovenko91} and generalize it to the case of
two SDWs. We consider the Hamiltonian
\begin{eqnarray}
{\cal H}_{\rm MF} &=& \sum _{K_x,k_y} 
\Bigl (\tilde \psi _+^\dagger (K_x,k_y), \,\,
\tilde \psi _-^\dagger (K_x,k_y) \Bigr ) \nonumber \\ && \times 
\left ( 
\begin{array}{lr}
v_FK_x  & \Delta e^{-i\phi (k_y)} \\ 
\Delta e^{i\phi (k_y)}
& -v_FK_x \\
\end{array}
\right )
\left (
\begin{array}{l}
\tilde \psi _+(K_x,k_y) \\ 
\tilde \psi _-(K_x,k_y)
\end{array} \right )
\end{eqnarray}
describing spinless electrons in presence of a charge-density wave at wave
vector $(2k_F,Q_y=0)$. Here $\Delta $ and $\phi (k_y)$ are real. 
$K_x=k_x\mp k_F$ for $\pm $ electrons. The spectrum is given by
$E_+=(v_F^2K_x^2+\Delta ^2)^{1/2}$ for the upper (empty) band and
$E_-=-(v_F^2K_x^2+\Delta ^2)^{1/2}$ for the lower (filled) band. The
wave functions are defined on the Brillouin zone torus $\vert k_x\vert <k_F$
and $0\leq k_y \leq 2\pi /b$. Consider now a wave function $\vert \psi
(k_x^{(0)},k_y)\rangle $ in the lower band. When $k_x$ is
changed along the closed line encircling the torus at fixed $k_y$, the wave
function changes and transforms into $e^{i\phi (k_y)}\vert \psi
(k_x^{(0)},k_y)\rangle $ when we return to the starting point. 
\cite{Yakovenko91,Kohmoto89} Noting that the second term in the rhs of
(\ref{Sxy}) vanishes (since the Hamiltonian is the same for $k_y=0$ and
$k_y=2\pi $), we obtain\cite{Yakovenko91}
\begin{equation}
\sigma _{xy} = \frac{e^2}{4\pi ^2} \lbrack \phi (2\pi /b)-\phi (0)\rbrack
.
\end{equation}
We now consider electrons with spin in presence of two SDWs (with zero
transverse wave vectors: $Q_y=0$): 
\begin{eqnarray}
{\cal H}_{\rm MF} &=& \sum _{K_x,k_y} 
\Bigl (\tilde \psi _{+\uparrow }^\dagger(K_x,k_y), \,\,
\tilde \psi _{-\downarrow }^\dagger(K_x,k_y) \Bigr ) \nonumber \\ && \times 
\left ( 
\begin{array}{lr}
v_FK_x  & \Delta _1e^{-i\phi _1(k_y)} \\ 
\Delta _1e^{i\phi _1(k_y)}
& -v_FK_x \\
\end{array}
\right ) 
\left (
\begin{array}{l}
\tilde \psi _{+\uparrow }(K_x,k_y) \\ 
\tilde \psi _{-\downarrow }(K_x,k_y)
\end{array} \right ) \nonumber \\ && + 
\sum _{K_x,k_y} 
\Bigl (\tilde \psi _{+\downarrow }^\dagger (K_x,k_y), \,\,
\tilde \psi _{-\uparrow }^\dagger (K_x,k_y) \Bigr ) \nonumber \\ && \times 
\left ( 
\begin{array}{lr}
v_FK_x  & \Delta _2e^{-i\phi _2(k_y)} \\ 
\Delta _2e^{i\phi _2(k_y)}
& -v_FK_x \\
\end{array}
\right ) 
\left (
\begin{array}{l}
\tilde \psi _{+\downarrow }(K_x,k_y) \\ 
\tilde \psi _{-\uparrow }(K_x,k_y)
\end{array} \right )  \nonumber \\ 
\label{ham6}
\end{eqnarray}
where $\Delta _1$, $\Delta _2$, $\phi _1$, and $\phi _2$  are real. 
We apply the same reasoning as
before to the wave function $\vert \psi _\uparrow (k_x^{(0)},k_y)\rangle $ of
the lower band. After the trip around the Brillouin zone, we
obtain the wave function $e^{i\phi _1(k_y)}\vert \psi _\downarrow
(k_x^{(0)},k_y)\rangle $. Repeating this procedure once more, we obtain the
wave function $e^{i\phi _1(k_y)+i\phi _2(k_y)}\vert \psi _\uparrow
(k_x^{(0)},k_y)\rangle $. This yields 
\begin{equation}
\sigma _{xy} = \frac{e^2}{4\pi } \lbrack \phi _1(2\pi /b)+ \phi _2(2\pi /b)
-\phi _1(0)-\phi _2(0) \rbrack .
\label{Sxy1} 
\end{equation}
It was shown in Ref.\ \onlinecite{Yakovenko91} that (\ref{Sxy1}) does not
depend on the value of the transverse wave vectors of the SDWs. Moreover, 
the global contribution of the gaps (if any) located below the Fermi level
vanishes. \cite{Yakovenko91} We are therefore now in
a position to calculate $\sigma _{xy}$ for the FISDW phases, using the 
Hamiltonian derived in the QLA (section \ref{sec:spec}).

\subsection{Sinusoidal waves}

Comparing (\ref{ham5}) and (\ref{ham6}), we deduce 
\begin{eqnarray}
\Delta _1e^{-i\phi _1(k_y)}&=& \Delta _2e^{-i\phi _2(k_y)} \nonumber \\
&=& -e^{iNk_yb} \vert \tilde \Delta _{+\uparrow }^{(N)} 
\vert - e^{-iNk_yb} \vert \tilde \Delta _{+\uparrow }^{(\bar N)} \vert .
\label{P12}
\end{eqnarray}
We have used $\vert \tilde \Delta _{+\uparrow }^{(pN)}\vert =\vert \tilde
\Delta _{-\uparrow }
^{(pN)}\vert $. When $k_y$ varies from $0$ to $2\pi /b$, the change $\phi _1
(2\pi /b)-\phi _1(0)=\phi _2(2\pi /b)-\phi _2(0)$ is determined by the term
with the largest amplitude in 
(\ref{P12}). Since $\vert \tilde \Delta _{+\uparrow }^{(\bar N)}/\tilde \Delta
_{+\uparrow }^{(N)}\vert =\vert \tilde \gamma \vert < 0.518$ in the
sinusoidal phase, we obtain
\begin{equation}
\phi _1(2\pi /b)+ \phi _2(2\pi /b) -\phi _1(0)-\phi _2(0) = -4\pi N.
\end{equation}
This yields 
\begin{equation}
\sigma _{xy} = -2N \frac{e^2}{h} ,
\label{QHEsin}
\end{equation}
where we have restored the dimensional Planck constant $h$. Thus, in the
sinusoidal phase, the QHE is determined by the SDW with the largest
amplitude.

\subsection{Helicoidal waves }

For helicoidal waves, we deduce from (\ref{ham4}) 
\begin{eqnarray}
\Delta _1 e^{-i\phi _1(k_y)} &=& -e^{iNk_yb} \tilde \Delta _{+\uparrow }^{(N)}
, \nonumber \\
\Delta _2 e^{-i\phi _2(k_y)} &=& -e^{-iNk_yb} \tilde \Delta _{-\uparrow }
^{(\bar N)} . 
\end{eqnarray} 
This yields $\phi _1(2\pi /b)-\phi _1(0)=-\lbrack \phi _2(2\pi /b)-\phi
_2(0)\rbrack $ so that  
\begin{equation}
\sigma _{xy}=0 .
\end{equation}
Thus, we come to the conclusion that the QHE vanishes in the helicoidal
phase.

\section{Magnetoelectric effect} 
\label{sec:mee} 
 
\begin{figure}
\epsfysize 5.cm 
\epsffile[100 375 430 580]{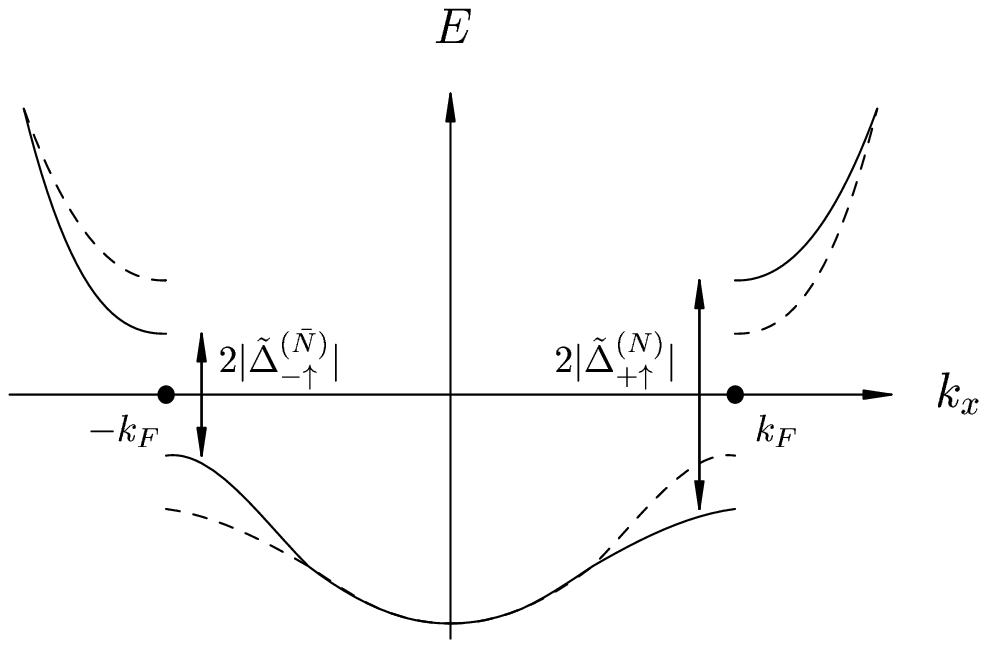}
\refstepcounter{figure} \label{Fig:spec2}
{\small FIG. \ref{Fig:spec2}.
Spectrum of electronic excitations in the helicoidal FISDW
phase. The solid (dashed) line corresponds to up (down) spins. Only the gaps
at the Fermi level are shown. For clarity, we have not shown the Zeeman
splitting. }
\end{figure}

A magnetoelectric effect may exist if time-reversal
and space-inversion symmetries are broken.\cite{Landau} 
Gor'kov and Sokol found the kinetic magnetoelectric effect
for a single helicoidal SDW.\cite{Gorkov87} The effect also
exists in the presence of two helicoidal SDWs of opposite chiralities,
provided their amplitudes are not equal.  An electric current $j_x$ along
the chains induces a uniform magnetization $\delta {\bf M}$ along the
vector ${\bf n}$ that characterizes the spin polarization of the
helicoidal SDWs. In our case, the vector {\bf n} is parallel to the
magnetic field {\bf H}, which is oriented along the $z$ axis, thus
\begin{equation}
\delta M_z\propto j_x.
\end{equation}  
Here $\delta M_z$ is the additional spin-magnetization density 
induced by $j_x$ in excess of the magnetization density $M_z$ induced by the
magnetic field without $j_x$.  The effect can be understood by
considering the spectrum of electronic excitations in the helicoidal
FISDW phase shown in Fig.\ \ref{Fig:spec2}.  The $+k_F$ electrons with
spin up and the $-k_F$ electrons with spin down have the energy gap
$|\tilde \Delta_{N,+}|$, whereas the $+k_F$ electrons with spin down
and the $-k_F$ electrons with spin up have the different energy gap
$|\tilde \Delta_{\bar N,-}|$. To produce a current $j_x$ along the
chains, electrons need to be transferred from $-k_F$ to $+k_F$ (we assume
that the electric field is weak enough so that the SDWs remain pinned by
impurities). For $|\tilde \Delta_{N,+}| \neq |\tilde
\Delta_{\bar N,-}|$ ($|\tilde \Delta_{N,+}| \neq |\tilde \Delta_{\bar N,-}|$
if $t_{4b}\neq 0$ and $N\ne 0$),
this redistribution of electrons affects up and down spins in a different
way, which results in a uniform magnetization $M_z$.
Denoting the deviation of the distribution 
function of electrons with spin $\sigma$ and momenta near
$\alpha k_F$ from the equilibrium one by $\delta n_{\alpha,\sigma}$, we have   
\begin{eqnarray}
j_x &=& ev_F \sum_\sigma (\delta n_{+,\sigma}-\delta n_{-,\sigma}),
\nonumber \\ 
\delta M_z &=& \frac{g\mu_B}{2} \sum_{\alpha=\pm}
(\delta n_{\alpha,\uparrow} - \delta n_{\alpha,\downarrow}).
\end{eqnarray}
Here we denote the electron gyromagnetic factor by $g$ (assumed to be equal
to 2 in the preceding 
sections). At low temperature, $T\ll |\tilde \Delta_{N,+}|, |\tilde
\Delta_{\bar N,-}|$, the electrons are excited solely above the
lowest energy gap ($\tilde \Delta_{\bar N,-}$ in Fig.\ \ref{Fig:spec2}). This
implies that $\delta n_{+,\uparrow }\simeq \delta n_{-,\downarrow }\simeq 0$
and 
\begin{equation}
\frac{\delta M_z}{j_x}\simeq -\frac{g\mu _B}{2ev_F}.
\label{mee1}
\end{equation}
Eq.\ (\ref{mee1}) can be rewritten as 
\begin{equation}
\frac{\delta M_z}{\mu _B}\simeq -\frac{gI}{2ev_FL_yL_z},
\end{equation}
where $I=j_xL_yL_z$ is the current passing through the sample of
cross-section  $L_yL_z$. 
For $I\sim 1 \mu{\rm A}$ (which is slightly below the critical current for
the depinning of the SDWs \cite{Balicas_th}), $L_yL_z\sim 1$ mm$^2$, $v_F\sim
3\times 10^{-5}$ m/s, and $g\simeq 2$, we obtain 
\begin{equation}
\frac{|\delta M_z|}{\mu _B}\sim 2\times 10^{-15}\,\, \AA^{-3}. 
\end{equation}
This should be
compared with the ground-state magnetization density 
\begin{equation}
\frac{|M_z|}{\mu _B}\sim 10^{-8}\,\, \AA^{-3},
\end{equation}
that we obtain from Ref.\ \onlinecite{Montambaux88}. Thus, we obtain 
$|\delta M_z|/|M_z|\sim 2\times 10^{-7}$.

The reentrant phases $N=0$ are somehow special since $|\tilde
\Delta_{+\uparrow }^{(0)}|=|\tilde \Delta_{-\uparrow }^{(\bar 0)}|$
independent of $t_{4b}$. Consequently, these phases do not exhibit the
magnetoelectric effect, although they are helicoidal.

\section{Coexistence between successive phases}
\label{sec:coex}

It has been shown by Lebed' that under certain circumstances, umklapp
processes can lead to the simultaneous existence of two successive
sinusoidal phases $N+\beta $ and $N$. \cite{Lebed90}
The system then evolves from the phase $N+\beta $ to the
phase $N$ via a region of the phase diagram where both the phases $N+\beta $
and $N$ exist. The transitions to the coexistence region are of second
order. 

In this section, we reconsider the problem of the coexistence between
two successive SDW phases, distinguishing between sinusoidal and helicoidal
waves. We analyze all possible cases (depending on the polarization of the
SDWs). 

The coexistence of two phases implies the simultaneous 
formation of four SDWs. Therefore Eqs.\ (\ref{pa}) and (\ref{paQLA}) should be
replaced by 
\bleq
\begin{eqnarray}
\tilde \Delta _{\alpha \sigma } (w_1,w_2) &=& \sum _{p=\pm } \sum _{\gamma
=0,\beta } 
\delta _{k_{2x},k_{1x}+\alpha Q_x^{(p(N+\gamma))}} 
(g_2 \Delta _{\alpha \sigma }^{(p(N+\gamma))}
+g_3 \Delta _{\bar \alpha \sigma }^{(\bar p(N+\gamma ))}) 
e^{-i\alpha Q_y^{(p(N+\gamma ))}b(l_1+l_2)/2} I_{\alpha (l_1-l_2)}
(Q_y^{(N+\gamma )}) ,
\nonumber \\ 
\tilde \Delta _{\alpha \sigma } (w_1,w_2) \Big \vert _{\rm QLA} &=& \sum
_{p=\pm }  
\sum _{\gamma =0,\beta } \delta _{k_{2x},k_{1x}+\alpha Q_x^{(p(N+\gamma ))}} 
\delta _{l_2,l_1-\alpha p(N+\gamma )} 
e^{-i\alpha Q_y^{(p(N+\gamma ))}b\lbrack 2l_1-\alpha p(N+\gamma )\rbrack /2} 
\tilde \Delta _{\alpha \sigma }^{(p(N+\gamma ))} .
\end{eqnarray}
The free energy becomes 
\begin{equation}
F_{N,N+\beta } = F_N + F_{N+\beta } + F^{\rm int}_{N,N+\beta } ,
\end{equation}
with the interacting part 
\begin{eqnarray}
F^{\rm int}_{N,N+\beta } &=& 2K \sum _\alpha \Bigl \lbrack 
\vert \tilde \Delta _{\alpha \uparrow }^{(N)} \vert ^2 +
\vert \tilde \Delta _{\alpha \uparrow }^{(\bar N)} \vert ^2 \Bigr \rbrack 
\Bigl \lbrack 
\vert \tilde \Delta _{\alpha \uparrow }^{(N+\beta )} \vert ^2 +
\vert \tilde \Delta _{\alpha \uparrow }^{(\overline{N+\beta })} \vert ^2 
\Bigr \rbrack 
\nonumber \\ && 
+2 K \cos \lbrack (N+\beta )Q_y^{(N)}b-NQ_y^{(N+\beta )}b \rbrack 
\sum _\alpha ( 
\tilde \Delta _{\alpha \uparrow }^{(N)}
\tilde \Delta _{\alpha \uparrow }^{(\bar N)} 
\tilde \Delta _{\alpha \uparrow }^{(N+\beta )*}
\tilde \Delta _{\alpha \uparrow }^{(\overline {N+\beta })*} + {\rm c.c.} ) 
\nonumber \\ && 
+ \delta _{N,0} \Bigl ( \sum _n \delta _{Q_y^{(0)}b,n\pi } \Bigr ) K
\sum _\alpha 
\Bigl \lbrack 2\sum _{p=\pm } \vert \tilde \Delta _{\alpha \uparrow
}^{(p\beta )}\vert ^2  
\tilde \Delta _{\alpha \uparrow }^{(0)}\tilde \Delta _{\alpha \uparrow
}^{(\bar 0)*} 
+\cos (\beta Q_y^{(0)}b) \lbrack (\tilde \Delta _{\alpha \uparrow }^{(0)})^2
+(\tilde \Delta _{\alpha \uparrow }^{(\bar 0)})^2 \rbrack    
\tilde \Delta _{\alpha \uparrow }^{(\beta )*}
\tilde \Delta _{\alpha \uparrow }^{(\bar \beta )*}
+ {\rm c.c.} \Bigr \rbrack .  \nonumber \\ && 
\label{Fint}
\end{eqnarray}

Without loss of generality, we have assumed that the phase $N+\beta $ is not
the last phase of the cascade (i.e., we do not consider the case $N+\beta
=0$ and $Q_y^{(N+\beta )}=\pi /b$). 
In order to study the possible coexistence of phases $N$ and $N+\beta $, we
consider a region in the phase diagram where the phase $N$ is  more stable
than the phase $N+\beta $ ($T_c^{(N)}>T_c^{(N+\beta )}$). Assuming that the
order 
parameters of the phase $N+\beta $ are infinitesimal, we derive an effective
free energy for this phase, from which we conclude about the coexistence of
the two phases. The same reasoning is applied to regions of the phase
diagram where $T_c^{(N)}<T_c^{(N+\beta )}$.

\subsection{Sinusoidal waves}

\subsubsection{$N\neq 0$}

Introducing the order parameters $u_\alpha ^{(N)}$, $v_\alpha ^{(N)}$,
$u_\alpha 
^{(N+\beta )}$, and $v_\alpha ^{(N+\beta )}$ defined in section
\ref{subsec:polarA}, 
and setting $v_\alpha ^{(N)}=v_\alpha ^{(N+\beta )}=0$, we rewrite
(\ref{Fint}) (for $N\neq 0$) as  
\begin{eqnarray}
F^{\rm int}_{N,N+\beta }&=& 2 K \sum _\alpha \Bigl \lbrack 
\cos ^2(\vartheta _N) \vert u^{(N)}_\alpha \vert ^2 +
\sin ^2(\vartheta _N) \vert u^{(N)}_{\bar \alpha }\vert ^2 \Bigr \rbrack 
\Bigl \lbrack 
\cos ^2(\vartheta _{N+\beta }) \vert u^{(N+\beta )}_\alpha \vert ^2 +
\sin ^2(\vartheta _{N+\beta }) \vert u^{(N+\beta )}_{\bar \alpha }\vert ^2
\Bigr \rbrack  \nonumber \\ &&
+K \cos \lbrack (N+\beta )Q_y^{(N)}b-NQ_y^{(N+\beta )}b \rbrack 
\sin (2\vartheta _N)\sin (2\vartheta _{N+\beta }) 
(u^{(N)}_+ u^{(N)}_- u^{(N+\beta )*}_+ u^{(N+\beta )*}_-+{\rm c.c.} )  .
\end{eqnarray}
For sinusoidal waves, we have $\vert u_+^{(N)}\vert = \vert u_-^{(N)}\vert $
and 
$\vert u_+^{(N+\beta )}\vert = \vert u_-^{(N+\beta )}\vert $.  
We consider a region of the phase diagram where $T_c^{(N)}>T_c^{(N+\beta )}$
and 
assume that $u_\alpha ^{(N+\beta )}$ is infinitesimal. To lowest order in
$u_\alpha ^{(N+\beta )}$, the effective free energy for the phase $N+\beta $
is  
\begin{eqnarray} 
F^{\rm eff}_{N+\beta }&=& 2\lambda _1^{(N+\beta )} \vert u_+ ^{(N+\beta
)}\vert ^2 +4K \vert u^{(N)}_+ \vert ^2 
 \vert u^{(N+\beta )}_+ \vert ^2 \nonumber \\ 
&& + K \cos \lbrack (N+\beta )Q_y^{(N)}b-NQ_y^{(N+\beta )}b \rbrack 
\sin (2\vartheta _N)\sin (2\vartheta _{N+\beta }) (u^{(N)}_+ u^{(N)}_-
u^{(N+\beta )*}_+ u^{(N+\beta )*}_-+{\rm c.c.} )  ,
\end{eqnarray} 
where $u^{(N)}_\alpha $ is not changed by the infinitesimal $u_\alpha
^{(N+\beta )}$ 
and is therefore given by (\ref{poS}) for $T\leq T_c^{(N)}$. The free
energy is minimal if the phase of $u_\alpha ^{(N+\beta )}$ is such that 
\begin{eqnarray}
\cos \lbrack (N+\beta )Q_y^{(N)}b-NQ_y^{(N+\beta )}b \rbrack 
\sin (2\vartheta _N)\sin (2\vartheta _{N+\beta }) 
u^{(N)}_+ u^{(N)}_- u^{(N+\beta )*}_+ u^{(N+\beta )*}_- && \nonumber \\ = - 
\vert \cos \lbrack (N+\beta )Q_y^{(N)}b-NQ_y^{(N+\beta )}b \rbrack 
\sin (2\vartheta _N)\sin (2\vartheta _{N+\beta }) \vert 
\vert u_+^{(N)} u_+^{(N+\beta )} \vert ^2 , &&
\end{eqnarray}
which yields 
\begin{equation}
F^{\rm eff}_{N+\beta } = 2\vert u_+^{(N+\beta )} \vert ^2 
\Bigl \lbrack \lambda _1^{(N+\beta )} +2 K \vert u^{(N)}_+ \vert ^2
-K \vert u^{(N)}_+ \vert ^2\vert \cos \lbrack (N+\beta
)Q_y^{(N)}b-NQ_y^{(N+\beta )}b \rbrack \sin (2\vartheta _N)\sin (2\vartheta
_{N+\beta }) \vert \Bigr \rbrack .
\end{equation}
\eleq
A second order phase transition to a phase where both $u_+^{(N)}$ and
$u_+^{(N+\beta )}$ are nonzero occurs if the coefficient of $\vert
u_+^{(N+\beta )} \vert ^2$  in $F^{\rm eff}_{N+\beta }$ becomes
negative. Since in all cases, $Q_y^{(N)},Q_y^{(N+\beta )}\sim \pi /b$ (see
section \ref{sec:imp}), we make the approximation $\vert 
\cos \lbrack (N+\beta )Q_y^{(N)}b-NQ_y^{(N+\beta )}b \rbrack \vert \simeq
1$. Using (see appendix \ref{app:lam})
\begin{equation}
\lambda _1^{(N)} \simeq \frac{N(0)}{2T_c^{(N)}} (T-T_c^{(N)}) ,
\end{equation}
we can write the coexistence condition for $T\leq T_c^{(N)}$ as
\begin{equation}
\frac{T-T_c^{(N+\beta )}}{T_c^{(N+\beta )}} \leq 
\frac{T-T_c^{(N)}}{T_c^{(N)}} x_+ ,
\label{cond1}
\end{equation}
where 
\begin{equation}
x_+ = \frac{2 - \vert \sin (2\vartheta _N)\sin (2\vartheta _{N+\beta }) \vert
}{1 + \frac{1}{2} \sin ^2(2\vartheta _N) } .
\end{equation}

For the coexistence in regions of the phase diagram where
$T_c^{(N)}<T_c^{(N+\beta )}$, we obtain by the same method the condition (for
$T\leq T_c^{(N+\beta )}$) 
\begin{equation}
\frac{T-T_c^{(N)}}{T_c^{(N)}} \leq 
\frac{T-T_c^{(N+\beta )}}{T_c^{(N+\beta )}} x_- ,
\label{cond2}
\end{equation}
where 
\begin{equation}
x_- = \frac{2 - \vert \sin (2\vartheta _N)\sin (2\vartheta _{N+\beta }) \vert
}{1 + \frac{1}{2} \sin ^2(2\vartheta _{N+\beta }) } .
\label{x-}
\end{equation}

$\vartheta _N$, $\vartheta _{N+\beta }$ are determined by $A_N$,
$A_{\bar N}$ and $B$, and therefore depend on temperature and magnetic
field. To verify 
that the region of the phase diagram defined by Eqs.\ (\ref{cond1}) and
(\ref{cond2}) exists, we
consider a point in the $(H,T)$-plane slightly below the intersection point
of the two curves $T_c^{(N)}$ and $T_c^{(N+\beta )}$. For this point
$T_c^{(N)} \simeq 
T_c^{(N+\beta )}$ and the coexistence conditions can be rewritten as 
\begin{equation}
(1-x) \frac{T-T_c^{(N)}}{T_c^{(N)}} \leq 0 ,
\end{equation}
where $x=x_-,x_+$. Since $T-T_c^{(N)}<0$, the coexistence is possible only if
$x_-<1$ or $x_+<1$. The existence of linearly polarized SDWs requires $\sin
^2(2\vartheta _N),\sin ^2(2\vartheta _{N+\beta })<2/3$ and therefore implies 
$x_-\vert _{\rm min}=x_+\vert _{\rm min}=1$. Consequently, there is no
coexistence between phases for sinusoidal waves. Thus, our results
invalidate Lebed's conclusion\cite{Lebed90} concerning the coexistence of
two successive sinusoidal phases. 

If  we discard the existence of helicoidal waves, then the
coexistence between phases may be possible depending on the geometry of the
Fermi surface. Let us take for instance $Q_y^{(N)}=Q_y^{(N+\beta )}=\pi /b$
and $t_{4b}=0$, which is the case considered by Lebed'. Then $\vartheta
_N=\vartheta _{N+\beta }=\pi /4$. The region of coexistence does exist and is
determined by 
\begin{equation}
\frac{3}{2} \frac{T-T_c^{(N+\beta)}}{T_c^{(N+\beta )}}
\leq \frac{T-T_c^{(N)}}{T_c^{(N)}} \leq \frac{2}{3}
\frac{T-T_c^{(N+\beta )}}{T_c^{(N+\beta )}} .
\label{coexLb}
\end{equation}
This is precisely the result obtained by Lebed'.\cite{Lebed90}

\subsubsection{$N=0$, $Q_y^{(0)}=\pi /b$}

We can study the coexistence of the last two phases of the cascade
following the same procedure. Using the result of section
\ref{subsec:polarN0}, we have
\begin{eqnarray}
F_{0,\beta }^{\rm int} &=& 8K\vert u\vert ^2\vert u_+^{(\beta )} \vert ^2 
\nonumber \\ &&
+2K \cos (\beta \pi ) \sin (2\vartheta _\beta )(u^2u_+^{(\beta )*}
u_-^{(\beta )*} + {\rm c.c.} ) .
\end{eqnarray}
Choosing the phases of  $u_+^{(\beta )} $ and $u_-^{(\beta )} $ in order to
minimize the free energy, we obtain
\begin{equation}
F_{0,\beta }^{\rm int} = 4K \lbrack 2- \vert \sin (2\vartheta _\beta
)\vert \rbrack \vert u\vert ^2\vert u_+^{(\beta )} \vert ^2 .
\end{equation}
We deduce that the coexistence region is determined by
\begin{equation}
\frac{1}{x_+'} \frac{T-T_c^{(\beta )} }{T_c^{(\beta )} } \leq 
\frac{T-T_c^{(0)}}{T_c^{(0)}} \le  x_-' 
\frac{T-T_c^{(\beta )} }{T_c^{(\beta )} } ,
\label{coex1}
\end{equation}
where 
\begin{eqnarray}
x_-' &=& \frac{2-\vert \sin (2\vartheta _\beta )\vert }{1+\frac{1}{2} \sin ^2(2
\vartheta _\beta )} , \nonumber \\ 
x_+' &=& 2-\vert \sin (2\vartheta _\beta )\vert .
\label{coex2}
\end{eqnarray}
The coexistence occurs if $x_+'<1$ or $x_-'<1$.
Since $\sin ^2(2\vartheta _\beta )<2/3$, $x_-'\vert _{\rm min}\simeq 0.89$. In
principle, the coexistence between the phases $N=\beta $ and $N=0$, $Q_y=\pi
/b$ is therefore 
possible in the region of the phase diagram where $T_c^{(\beta )} >T_c^{(0)}$
provided that $x_-'<1$. If $\beta =1$, then our numerical calculations show
that $\sin ^2(2\vartheta _\beta )\ll 2/3$ (the latter inequality holds for all
phases with $N$ odd). Therefore, the coexistence region does not exist if the
phase $N=0$ is preceded by the phase $\beta =1$. In the 
case where the  phase $N=0$ is preceded by the phase $\beta
=2$, the coexistence region exists provided that $x_-'<1$. This situation
however requires strong umklapp scattering and is therefore quite unlikely.
(Note that for $\sin ^2(2\vartheta _\beta )=1$, Eq.\ (\ref{coex1}) agrees with
Lebed's results.\cite{Lebed90})

\subsubsection{$N+\beta =0$, $Q_y^{(N+\beta )}\neq \pi /b$}
\label{subsubsec:coexN0}

The coexistence between the phase $N$ and the reentrant phase $N+\beta =0$,
$Q_y^{(N+\beta )}\neq \pi /b$ can be studied by setting $\beta =-N$. This
case is special since the phase $N+\beta =0$, $Q_y^{(N+\beta )}\neq \pi /b$
is always 
helicoidal when $g_3\neq 0$ (see section \ref{sec:polar}). Using $\vert
u^{(N)}_+\vert =\vert u^{(N)}_-\vert $ and $u^{(0)}_-=0$, we have 
\begin{equation}
F^{\rm int}_{N,0}= 2K\vert u^{(0)}_+u^{(N)}_+ \vert ^2 .
\end{equation}
In the region $T_c^{(N)}<T_c^{(0)}$, we find the effective free
energy for the phase $N$ (for $T<T_c^{(0)}$)
\begin{equation}
F^{\rm eff}_N = \vert u_+^{(N)} \vert ^2 N(0)  
\left \lbrack \frac{T-T_c^{(N)}}{T_c^{(N)}} -2 
\frac{T-T_c^{(0)}}{T_c^{(0)}} \right \rbrack .  
\end{equation} 
The coefficient of $\vert u_+^{(N)}\vert ^2$ is always positive so that
there is no phase coexistence. In the region $T_c^{(0)}<T_c^{(N)}$, we
find the effective free energy (for $T<T_c^{(N)}$)
\begin{eqnarray}
F_0^{\rm eff}&=&\vert u_+^{(0)} \vert ^2 \frac{N(0)}{2}   
\Biggl \lbrack \frac{T-T_c^{(0)}}{T_c^{(0)}} \nonumber \\ && 
- \frac{2}{1+\frac{1}{2}\sin
^2(2\vartheta _N)} \frac{T-T_c^{(N)}}{T_c^{(N)}} \Biggr \rbrack ,  
\end{eqnarray} 
which shows that there is no coexistence.

\subsection{Helicoidal waves}

\subsubsection{$N\neq 0$, or $N=0$ and $Q_y^{(0)}\neq \pi /b$} 
\label{subsubsec:coexN0a} 

First we study the coexistence between the two helicoidal phases $N$ and
$N+\beta $. For simplicity, we consider only the case $\vartheta _N
=\vartheta _{N+\beta }=\pi 
/4$. For helicoidal waves, we have $u_-^{(N)}=0$ and $u_-^{(N+\beta )}=0$
(we could also choose $u_+^{(N+\beta )}=0$, this would not change the
result). Eq.\ (\ref{Fint}) then yields
\begin{equation}
F_{N,N+\beta }^{\rm int}= K\vert u_+^{(N)} u_+^{(N+\beta )}\vert ^2 .
\end{equation}
In the region $T_c^{(N)}>T_c^{(N+\beta )}$, we find the effective free
energy for the phase $N+\beta $ (for $T<T_c^{(N)}$) 
\begin{equation}
F_{N+\beta }^{\rm eff}= \vert u_+^{(N+\beta )}\vert ^2 \frac{N(0)}{2} 
\left \lbrack \frac{T-T_c^{(N+\beta )}}{T_c^{(N+\beta )}} -2 
\frac{T-T_c^{(N)}}{T_c^{(N)}} \right \rbrack .  
\end{equation} 
The coefficient of $\vert u_+^{(N+\beta )} \vert ^2$ is always positive so
that there is no phase coexistence. In the region $T_c^{(N)}<T_c^{(N+\beta
)}$, we find the effective free energy (for $T<T_c^{(N+\beta )}$)
\begin{equation}
F_N^{\rm eff}= \vert u_+^{(N)} \vert ^2 \frac{N(0)}{2} 
\left \lbrack \frac{T-T_c^{(N)}}{T_c^{(N)}} -2 
\frac{T-T_c^{(N+\beta )}}{T_c^{(N+\beta )}} \right \rbrack ,  
\end{equation} 
which shows that there is no coexistence.
 
\subsubsection{$N=0$, $Q_y=\pi /b$} 

For the coexistence between the helicoidal phase $N=\beta $ and the
sinusoidal phase $N=0$, $Q_y=\pi /b$ terminating the cascade, we have
\begin{equation}
F_{0,\beta }^{\rm int}= 4K\vert u\vert ^2
\vert u_+^{(\beta )} \vert ^2 ,
\end{equation}
This yields (for $T<T_c^{(0)}$) 
\begin{equation}
F_\beta ^{\rm eff}= \vert u_+^{(\beta )} \vert ^2 \frac{N(0)}{2} 
\left \lbrack \frac{T-T_c^{(\beta )} }{T_c^{(\beta )} } -2 
\frac{T-T_c^{(0)}}{T_c^{(0)}} \right \rbrack   
\end{equation} 
in the region $T_c^{(0)}>T_c^{(\beta )} $, and (for $T<T_c^{(\beta )} $)  
\begin{equation}
F_0^{\rm eff}= 4\vert u\vert ^2 \frac{N(0)}{2} 
\left \lbrack \frac{T-T_c^{(0)}}{T_c^{(0)}} -2 
\frac{T-T_c^{(\beta )} }{T_c^{(\beta )} } \right \rbrack   
\end{equation} 
in the region $T_c^{(0)}<T_c^{(\beta )}$. The coexistence is therefore not
possible.

\section{Summary and conclusion}
\label{sec:sum}

Our main results can be summarized as follows:

In presence of umklapp processes, the instability of the metallic phase at
the temperature $T_c^{(N)}$ corresponds to the formation of two SDWs, with
quantized longitudinal wave vectors $Q_x^{(N)}=2k_F+NG$ and $Q_x^{(\bar
N)}=2k_F-NG$. For very weak umklapp scattering, both SDWs are incommensurate
in the transverse direction ($Q_y^{(N)}=-Q_y^{(\bar N)}\ne \pi /b$) except
when $N=0$. If we
label each phase by the integer $N$ corresponding to the SDW with the
largest amplitude, we have sgn($N$)=sgn($t_{2b}$). The amplitude of the SDW
at wave vector $Q_x^{(\bar N)}$ is vanishingly small. 
The quantum Hall conductivity is determined by the SDW
with the largest amplitude, i.e., $\sigma _{xy}=-2Ne^2/h$. 

For even $N$, there exists a critical value of $g_3$ (typically $g_3/g_2$ of
the order of a few percent) above which the system
prefers to form two transversely commensurate SDWs ($Q_y^{(N)}=-Q_y^{(\bar
N)}= \pi /b$). When $t_{4b}>0$, the SDW with the largest amplitude is then
determined by sgn($N$)=$-$sgn($t_{2b}$): the QHE changes sign. The two
SDWs have comparable amplitudes when $t_{4b}$ is not too large
($t_{4b}\lesssim 1.5$ K). When the umklapp scattering strength
increases, the first negative phase to appear is the phase $N=-2$. 

Umklapp scattering also tends to suppress the phases with an odd $N$ and
produce some reentrances of the phase $N=0$ within the cascade. Unlike 
the last phase of the cascade, the reentrant phases $N=0$ are incommensurate
($Q_y\ne \pi /b$). 

The negative phases are likely to become helicoidal when the umklapp
scattering strength is further increased. Experimentally, this situation
could be achieved by decreasing pressure. The appearance of these helicoidal
phases is entirely controlled by $t_{4b}$. The QHE vanishes in the
helicoidal phases, but a magnetoelectric effect appears. These two
characteristic properties may be utilized to detect the helicoidal phases
experimentally. The reentrant phases $N=0$ are always helicoidal but do not
exhibit the magnetoelectric effect. 

In the sinusoidal phases, umklapp processes modulate the gap on the
Fermi surface as a function of $k_y$. When $\vert \Delta (k_y)\vert _{\rm
min}/\vert \Delta (k_y)\vert _{\rm max} \simeq 0.32$, the sinusoidal phase
becomes unstable against the formation of a helicoidal phase. 

The conclusion of Lebed' \cite{Lebed90} that, in the presence of umklapp 
scattering, adjacent FISDW phases are separated by two second-order phase
transitions and an intermediate phase with coexistence of four SDWs is
incorrect, because he did not consider helicoidal SDWs.

In conclusion, the consideration of umklapp scattering naturally explains
the appearance of negative FISDW phases in quasi-one-dimensional organic
conductors. These phases are characterized not only by a sign reversal of
the QHE, but also but the simultaneous presence of two SDWs with comparable
amplitudes (provided that $t_{4b}$ is not too large). This leads to the
possible stabilization of helicoidal phases. But even in the sinusoidal phases
(which are the ones that have been observed up to now), we expect the
presence of two SDWs to give rise to new physical properties.

\section*{Acknowledgments}

This work was partially supported by the NSF under Grant
DMR--9417451 and by the David and Lucile Packard Foundation.

\bleq

\appendix

\section{}
\label{app:pa}

We calculate in this appendix the pairing amplitudes 
\begin{eqnarray}
\tilde \Delta _{\alpha \sigma } (w_1,w_2) &=& \int d^2r \, 
\phi _{w_1}^{\bar \alpha *}({\bf r}) \phi _{w_2}^\alpha ({\bf r})
\tilde \Delta _{\alpha \sigma } ({\bf r}) \nonumber \\
&=& \sum _{p=\pm } \int d^2r \, 
\phi _{w_1}^{\bar \alpha *}({\bf r}) \phi _{w_2}^\alpha ({\bf r})
\Bigl (g_2 \Delta _{\alpha \sigma }^{(pN)} e^{-i\alpha {\bf Q}_{pN}\cdot
{\bf r}}  
+g_3 \Delta _{\bar \alpha \sigma }^{(pN)} e^{-i\alpha 4k_Fx+i\alpha {\bf
Q}_{pN}\cdot {\bf r}} \Bigr ) .
\end{eqnarray}
Using $-4k_F+Q_x^{(pN)}=-Q_x^{(\bar pN)}$ and $Q_y^{(pN)}=-Q_y^{(\bar pN)}$, we
have  
\begin{equation}
\tilde \Delta _{\alpha \sigma } (w_1,w_2) = 
\sum _{p=\pm } \left ( g_2 \Delta _{\alpha \sigma }^{(pN)}+g_3 \Delta _{\bar
\alpha 
\sigma }^{(\bar pN)} \right )
\int d^2r
\phi _{w_1}^{\bar \alpha *}({\bf r}) \phi _{w_2}^\alpha ({\bf r}) 
e^{-i\alpha {\bf Q}_{pN}\cdot {\bf r}} .
\end{equation}
Using (\ref{Idef}), we have 
\begin{eqnarray}
\int d^2r \phi _{w_1}^{\bar \alpha *}({\bf r}) \phi _{w_2}^\alpha ({\bf r}) 
e^{-i\alpha {\bf Q}_{pN}\cdot {\bf r}} &=& \frac{1}{L_x} \int dx\, 
e^{ix(-k_{1x}+k_{2x}-\alpha Q_x^{(pN)})} \sum _m f_{l_1-m}^{\bar \alpha }
f_{l_2-m}^\alpha e^{-i\alpha Q_y^{(pN)}bm} \nonumber \\
&=& \delta _{k_{2x},k_{1x}+\alpha Q_x^{(pN)}} 
e^{-i\alpha Q_yb(l_1+l_2)/2} 
I_{\alpha (l_1-l_2)} (\alpha Q_y^{(pN)}).
\label{app1}
\end{eqnarray}
Since $I_n(q_y)=I_n(-q_y)$, (\ref{app1}) leads to (\ref{pa}). 

\section{}
\label{app:fe}

In this appendix, we give the main steps in the calculation of the fourth
order term of the free energy leading to (\ref{F4}) and (\ref{Fint}). We
consider the case where there are simultaneously four SDWs at wave vectors
${\bf Q}_N$, ${\bf Q}_{\bar N}$, ${\bf Q}_{N+\beta }$, ${\bf Q}_{\overline
{N+\beta }}$:
\begin{eqnarray}
F^{(4)}_N &=& \frac{K}{2} \sum _\alpha \sum _{p_1,\cdots,p_4=\pm }
\sum _{\gamma _1,\cdots,\gamma _4=0,\beta }
\tilde \Delta _{\alpha \uparrow }^{(p_1(N+\gamma _1))}
\tilde \Delta _{\alpha \uparrow }^{(p_2(N+\gamma _2))*}
\tilde \Delta _{\alpha \uparrow }^{(p_3(N+\gamma _3))}
\tilde \Delta _{\alpha \uparrow }^{(p_4(N+\gamma _4))*} \nonumber \\ && \times
\exp \Bigl \lbrace i(b/2) \Bigl \lbrack p_1(N+\gamma _1)Q_y^{(p_1(N+\gamma
_1))}  
- \lbrack 2p_1(N+\gamma _1)-p_2(N+\gamma _2)\rbrack Q_y^{(p_2(N+\gamma
_2))} \nonumber \\ && 
+\lbrack 2p_1(N+\gamma _1)-2p_2(N+\gamma _2)+p_3(N+\gamma _3)\rbrack 
Q_y^{(p_3(N+\gamma _3))}-p_4(N+\gamma _4) Q_y^{(p_4(N+\gamma _4))}\Bigr
\rbrack \Bigr \rbrace \nonumber \\ && \times
\delta _{(p_1-p_2+p_3-p_4)N+p_1\gamma _1-p_2\gamma _2+p_3\gamma _3-p_4\gamma
_4,0} \sum _n \delta _{Q_y^{(p_1(N+\gamma _1))}-Q_y^{(p_2(N+\gamma _2))}+
Q_y^{(p_3(N+\gamma _3))}-Q_y^{(p_4(N+\gamma _4))},n2\pi/b} \,.
\label{app2}
\end{eqnarray}
We write $F^{(4)}_N=F^{(4)}_N\vert _4+F^{(4)}_N\vert _3+F^{(4)}_N\vert _2+
F^{(4)}_N\vert _1$. $F^{(4)}_N\vert _4$ corresponds to the case where all
the $(p_i,\gamma _i)$ in (\ref{app2}) are identical ($p_1=p_2=\cdots $,
$\gamma _1=\gamma _2=\cdots $). $F^{(4)}_N\vert _3$ corresponds to the case
where three of the $(p_i,\gamma _i)$ are equal and different from the fourth
one... $F^{(4)}_N\vert _1$ corresponds to the case where all the
$(p_i,\gamma _i)$ are different. Skipping the details of the calculation, we
only give the final result for $F^{(4)}_N\vert _i$: 
\begin{eqnarray}
F^{(4)}_N\vert _4 &=& \frac{K}{2} \sum _{\alpha ,p,\gamma } 
\vert \tilde \Delta _{\alpha \uparrow }^{(p(N+\gamma ))} \vert ^4 ,
\nonumber \\ 
F^{(4)}_N\vert _3 &=& 
\delta _{N,0} \Bigl ( \sum _n \delta _{Q_y^{(0)}b,n\pi } \Bigr ) K
\sum _\alpha \left \lbrack \tilde \Delta _{\alpha \uparrow }^{(0)}
\vert \tilde \Delta _{\alpha \uparrow }^{(\bar 0)} \vert ^2 
\tilde \Delta _{\alpha \uparrow }^{(\bar 0)*}
+ \tilde \Delta _{\alpha \uparrow }^{(\bar 0)}
\vert \tilde \Delta _{\alpha \uparrow }^{(0)} \vert ^2 
\tilde \Delta _{\alpha \uparrow }^{(0)*} + {\rm c.c.} \right \rbrack
\nonumber \\ 
F^{(4)}_N\vert _2 &=& K \sum _\alpha \sum _{p\gamma \neq p'\gamma '} 
\vert \tilde \Delta _{\alpha \uparrow }^{(p(N+\gamma ))} 
\tilde \Delta _{\alpha \uparrow }^{(p'(N+\gamma '))} \vert ^2 
\nonumber \\ && +  \delta _{N,0} \Bigl ( \sum _n \delta _{Q_y^{(0)}b,n\pi }
\Bigr ) K \sum _\alpha \Biggl \lbrack 2\sum _p \vert \tilde \Delta
_{\alpha \uparrow }^{(p\beta )}\vert ^2 \tilde \Delta _{\alpha \uparrow
}^{(0)}  
  \tilde \Delta _{\alpha \uparrow }^{(\bar 0)*}
+\frac{1}{2}(\tilde \Delta _{\alpha \uparrow }^{(0)}
\tilde \Delta _{\alpha \uparrow }^{(\bar 0)*})^2
\nonumber \\ && 
+\cos (\beta Q_y^{(0)}b) \lbrack (\tilde \Delta _{\alpha \uparrow }^{(0)})^2 
+(\tilde \Delta _{\alpha \uparrow }^{(\bar 0)})^2 \rbrack   
\tilde \Delta _{\alpha \uparrow }^{(\beta )*}
\tilde \Delta _{\alpha \uparrow }^{(\bar \beta )*}
+{\rm c.c.} \Biggr \rbrack \nonumber \\ 
F^{(4)}_N\vert _1 &=& 
2 K \cos \lbrack (N+\beta )Q_y^{(N)}b-NQ_y^{(N+\beta )}b \rbrack 
\sum _\alpha ( 
\tilde \Delta _{\alpha \uparrow }^{(N)}
\tilde \Delta _{\alpha \uparrow }^{(\bar N)} 
\tilde \Delta _{\alpha \uparrow }^{(N+\beta )*}
\tilde \Delta _{\alpha \uparrow }^{(\overline {N+\beta })*} + {\rm c.c.} ) 
\end{eqnarray}

\section{}
\label{app:lam}

In this appendix, we calculate $\lambda _1^{(N)}$ for $T\simeq
T_c^{(N)}$. Since 
$\lambda _1^{(N)}$ vanishes for $T=T_c^{(N)}$, we have 
\begin{eqnarray}
\lambda _1^{(N)} &\simeq &(T-T_c^{(N)}) \frac{\partial \lambda
_1^{(N)}}{\partial T} 
\Biggr \vert _{T_c^{(N)}} \nonumber \\ &\simeq & 
(T-T_c^{(N)}) \frac{\partial }{\partial T} \Biggl ( \frac{A_N+A_{\bar N}}{2}
+\frac{1}{2} {\rm sgn} (A_N-A_{\bar N})\lbrack (A_N-A_{\bar
N})^2+4B^2\rbrack ^{1/2} \Biggr ) _{T_c^{(N)}} . \nonumber \\
\end{eqnarray}
From (\ref{a12b}) we deduce $\partial B/\partial T=0$ and 
\begin{equation}
A_N= A_N\vert _{T_c^{(N)}} - \frac{T-T_c^{(N)}}{I_N^2} \frac{\partial \chi
_+^{(0)}({\bf Q}_N) } 
{\partial T} \Biggr \vert _{T_c^{(N)}} ,
\end{equation}
with
\begin{equation}
\frac{\partial \chi ^{(0)}_+({\bf Q}_N)}{\partial T} = \sum _n
I_n^2(Q_y^{(N)}) 
\frac{\partial \chi ^{\rm 1D}_+(2k_F+(N-n)G)}{\partial T}  .
\end{equation}
Eq.\ (\ref{chi1D}) yields 
\begin{equation}
\frac{\partial \chi ^{\rm 1D}_+(2k_F-nG)}{\partial T}= \frac{N(0)}{2} 
\Biggl \lbrace -\frac{1}{T}
+{\rm Re} \left \lbrack \frac{in\omega _c}{4\pi
T^2} \Psi '\left (\frac{1}{2}+\frac{in\omega _c}{4\pi T} \right ) \right
\rbrack \Biggr \rbrace , 
\end{equation}
where $\Psi '$ is the derivative of the digamma function. Using $\Psi
'(z)\simeq 1/z+1/2z^2$ for $\vert z\vert \gg 1$, we obtain for $\omega _c\gg
T$
\begin{equation}
\frac{\partial \chi ^{\rm 1D}_+(2k_F-nG)}{\partial T}= -\frac{N(0)}{2T}
\delta _{n,0} +O(T/\omega _c^2) .
\end{equation}
To leading order in $T/\omega _c$, we therefore have
\begin{equation}
\frac{\partial \chi ^{(0)}_+({\bf Q}_N)}{\partial T} = - I_N^2
\frac{N(0)}{2T},  
\end{equation}
from which we deduce
\begin{equation}
A_N= A_N\vert _{T_c^{(N)}} +\frac{N(0)}{2T_c^{(N)}} (T-T_c^{(N)}) .
\end{equation}
In the same way, we can show that 
\begin{equation}
A_{\bar N}= A_{\bar N}\vert _{T_c^{(N)}}
+\frac{N(0)}{2T_c^{(N)}}(T-T_c^{(N)}) . 
\end{equation}
This yields
\begin{equation}
\lambda _1^{(N)} \simeq \frac{N(0)}{2T_c^{(N)}} (T-T_c^{(N)}) .  
\end{equation}

\eleq

\ecols

\begin{references}
\bibitem{Lebed91a} The fact that umklapp processes can make the electronic 
spectrum gapless in the sinusoidal phase has also been noticed by
Lebed' in Ref.\ \onlinecite{Lebed91}.
\end{references}
\end{document}